\documentstyle[prb,aps,multicol,epsfig]{revtex}

\begin{document}

\draft

%----------------------------------------------------
\title{Effect of the spin-orbit interaction \\
on the band structure and conductance of quasi-one-dimensional systems}
%----------------------------------------------------

%----------------------------------------------------
\author{A.~V.~Moroz and C.~H.~W.~Barnes}

\address{Cavendish Laboratory,
     University of Cambridge, \\
     Madingley Road,
     Cambridge CB3 0HE,
     United Kingdom}
%----------------------------------------------------

\date{\today}

\maketitle

%----------------------------------------------------
\begin{abstract}

We discuss the effect of the spin-orbit interaction on the band structure,
wave functions and low temperature conductance of long
quasi-one-dimensional electron systems patterned in two-dimensional
electron gases (2DEG). Our model for these systems consists of a linear
(Rashba) potential confinement in the direction perpendicular to the 2DEG
and a parabolic confinement transverse to the 2DEG. We find that these two
terms can significantly affect the band structure introducing a wave vector
dependence to subband energies, producing additional subband minima and
inducing anticrossings between subbands. We discuss the origin of these
effects in the symmetries of the subband wave functions.

\end{abstract}
%------------------------------------------------------------------------


\begin{multicols}{2}
\narrowtext
%%%%%%%%%%%%%%%%%%%%%%%%%%%%% SECTION %%%%%%%%%%%%%%%%%%%%%%%%%%%%%%%%%%%
\section{Introduction}
\label{sec-intro}
%%%%%%%%%%%%%%%%%%%%%%%%%%%%%%%%%%%%%%%%%%%%%%%%%%%%%%%%%%%%%%%%%%%%%%%%%

In 1986, the first experimental realisation of a quasi-one-dimensional
electron system (Q1DES) in a dynamically confined two-dimensional electron
gas (2DEG) was achieved~\cite{Thornton}. Since then there has been an
extensive theoretical and experimental effort into understanding their
basic properties (see, e.g., review~\cite{BvH} and book~\cite{Kelly} and
references therein). The interest in these systems stems from two facts.
First, the effective transverse size of a Q1DES can be easily controlled
and made remarkably small, down to the de Broglie wavelength of an
electron. This makes it possible to realise experimental systems which
have an arbitrary number of occupied transverse modes. Second, the high
purity of 2DEGs grown by molecular beam epitaxy enables the almost
collisionless motion of an electron through an experimental Q1DES. The
coexistence of these two factors has made Q1DES unique objects for the
investigation of transport phenomena yielding, in particular, the
observation of the ballistic quantisation of
conductance~\cite{vanWees,Wharam} and the so called 0.7 conductance
structure~\cite{Thomas1,Kristensen,Thomas2}.

The process of electron transmission through a Q1DES involves the
redistribution of incoming electron flux among its discrete eigenstates
followed by adiabatic transport through them. Therefore the determination
of the electron eigenstates of a Q1DES is an integral and very important
part of the more general quantum transport problem. This statement is
especially relevant to the ballistic transport regime where the {\it total}
conductance of a system is {\it completely defined} by the number of
propagating electron modes which in turn can be uniquely calculated from
the energy spectrum and the Fermi energy~\cite{BvH,Kelly,Datta}.

Clearly, the energy spectrum of electrons crucially depends on the
effective geometry of a Q1DES as well as on external and internal fields
acting on them. Among the possible internal forces, one of the least
understood examples is the interaction between orbital and spin degrees of
freedom of an electron: the {\it spin-orbit interaction}, also referred to
as the spin-orbit coupling. Although this interaction has an essentially
relativistic nature (see, e.g., Refs.~\cite{Thankappan,Landau,Fisher}), it
nevertheless can give rise to an observable modification of semiconductor
band structure
\cite{Stormer,Stein,Das2,Das,Dresselhaus2,Jusserand,Goldoni,Chen,Nitta}.

A quite general theoretical approach to the description of the SO
interaction is to use the following Hamiltonian~\cite{Thankappan,Landau}
which stems directly from the quadratic in $v/c$ expansion of the Dirac
equation:

\begin{equation}
\hat H_{SO} = \frac{\hbar}{(2M_0c)^2}\nabla V({\bf r})
\left(\hat{\bbox{\sigma}}\times\hat{\bf p}\right).
\label{H_SO}
\end{equation}
Here $M_0$ is the free electron mass, $\hat{\bf p}$ is the momentum
operator, $e$ is the elementary charge,
$\hat{\bbox{\sigma}}=\left\{\sigma_x,\sigma_y,\sigma_z\right\}$ is the
vector of the Pauli matrices, $V({\bf r})$ is the potential energy of the
particle, and $\nabla$ stands for the spatial gradient. The convenience (or
universality) of the Hamiltonian (\ref{H_SO}) is that it does not restrict
one to any particular form (model) of the potential $V({\bf r})$, but
allows freedom of choice depending on the nature and the symmetry of forces
present in a given medium. Its form is such that it can remove the spin
degeneracy in electron band structure whilst not actually producing an
overall spin polarization.

In a bulk (3D) crystalline environment, the energy $V({\bf r})$ arises
exclusively from the periodic (microscopic) crystal potential. Most
multicomponent semiconductors have either zincblende (GaAs and most III-V
compounds) or wurtzite (II-VI compounds) lattice structure, both of which
lack inversion symmetry. Dresselhaus~\cite{Dresselhaus} has shown that
this property eventually leads to a SO-induced splitting of the conduction
band into two subbands. The magnitude of the splitting is proportional to
the cube of the electron wave number $k$.

In MOSFETs and heterostructures the host crystals cannot be treated as
ideal 3D systems, because the crystal symmetry is broken at the device
interface where 2D electron or hole gases are dynamically confined in a
quantum well. The reduction of the effective dimensionality lowers the
symmetry of the underlying crystals and results in an additional (linear in
$k$) term in the Dresselhaus splitting. Moreover, if the quantum well is
sufficiently narrow, then the linear contribution is
dominant~\cite{Jusserand,Luo,Eppenga,Malshukov} and, e.g., may reach $\sim
0.3$ -- $0.4$ meV in 180-\AA-thick modulation-doped GaAs
wells~\cite{Jusserand}. Theoretical arguments~\cite{Edelstein} suggest that
this can also be true in strained III-V crystals and hexagonal II-VI
compounds.

Along with the {\it microscopic} crystal forces, there is another source
of the potential energy $V({\bf r})$ in 2D systems. It is caused by the
interface electric field that accompanies the quantum-well
asymmetry~\cite{Kelly,Ando} and is directed along the normal to the
device plane. Since the typical well width is $\sim 1$ -- $10$ nm, the
interface potential turns out to be slowly varying on the scale of the
lattice parameter and can be considered {\it macroscopic}, as opposed to
the rapidly oscillating atomic field. The mechanism of the SO interaction
originating from the interface field was first introduced by Rashba in
Ref.~\cite{Rashba}. It also manifests itself as a linear in $k$
splitting~\cite{Bychkov} of the 2D bandstructure. In a variety of systems
including Si-MOSFETs~\cite{Dorozhkin}, InAs/GaSb~\cite{Luo} and
AlSb/InAs/AlSb~\cite{Chen} quantum wells, InGaAs/InAlAs
heterostructures~\cite{Das,Nitta}, GaAs electron gases~\cite{Hassenkam} it
can be made to dominate the Dresselhaus terms indicating the significance
of macroscopic potentials in producing observable SO effects in
low-dimensional systems.

In modern nanotechnology there exists a number of methods for creating
Q1DES from 2DEGs: the split-gate technique; wet and dry etching; and
cleaved edge overgrowth~\cite{Yacoby}. In essence, they  all exploit the
confinement of the lateral (in-plane) motion of electrons (holes) by some
transverse potential. Any such potential must be essentially non-uniform in
space in order to force the charged particles to remain within a confined
area. The spatial variation scale of the confining potential crucially
depends on the particular fabrication method and varies over a wide range:
$\sim 10$ -- $1000$ nm. Thus, in Q1DES one finds another example of
macroscopic potentials, viz the lateral confining potential, which is
absent in higher-dimensional structures. The spatial non-uniformity of the
confining potential gives rise to an additional (in-plane) electric field
in the system. If the confinement is sufficiently strong (narrow and deep),
then this field may not be negligible in comparison with the
interface-induced (Rashba) field. Moreover, in nearly square (i.e.
symmetric) quantum wells where the Rashba field is essentially
suppressed~\cite{Chen,Hassenkam}, the in-plane (``confining'') electric
field is likely to be dominant. This suggests the possible importance of
the lateral confining potential to the SO Hamiltonian (\ref{H_SO}) in
Q1DES. We are not aware of any experimental evidence or measurement of the
strength of the SO coupling resulting from such a confining potential and
cannot therefore quote a grounded estimate for the corresponding energy
modification. Nevertheless, the above arguments seem sufficiently strong to
indicate a possible new mechanism for the SO interaction in Q1DES and point
towards new transport effects. The existence of an additional (and easily
controllable) source of the SO coupling could catalyse experiments on
quantum-wire based devices that exploit both the charge and spin of an
electron, e.g. spin transistors~\cite{DD,Kane} and active spin
polarisers~\cite{Fasol}.

Our belief in the importance of the SO interaction in semiconductor Q1DES
is strongly supported by the fact that observable manifestations of a
SO-induced energy splitting have already been found
experimentally~\cite{Kveder} in another type of Q1DES, viz in electron
gases trapped by dislocational defects in silicon crystals. The source of
the SO coupling in these systems is an electric field perpendicular to
dislocations and the SO-related energy splitting is linear in the wave
number $k$, which makes a physical analogy between the electron gas on
dislocations and semiconductor Q1DES quite close. The SO interaction
manifests itself in spin-dependent electron conductivity along dislocations
which can be ascribed to combined spin resonance~\cite{Rashba,Rashba2} of
electrons corresponding to transitions between SO-split energy levels.

In this paper we analyse theoretically the effect of the spin-orbit
interaction on the energy spectrum and conductance of a long Q1DES. As a
model for the Q1DES, we use a strictly 2D electron gas subject to a
transverse electrostatic confining potential. To decide on a reasonable
shape of the confining potential, we assume a sufficiently small effective
width ($\lesssim 300$ nm) for the Q1DES and a low electron concentration
($\lesssim 10^{11}$ ${\rm cm}^{-2}$). Combined together, these two factors
prevent the confining electric field from being significantly screened by
the electron gas. Under these conditions the confining potential can be
accurately approximated by a parabolic
potential~\cite{Laux,Drexler,Kardynal}. This conclusion is very favourable
to our problem because exact analytical expressions for the energy spectrum
and wave functions of a 2D electron gas in a parabolic potential are well
known and provide us with a good zero approximation in dealing with the SO
coupling.

We include the SO interaction via the Hamiltonian (\ref{H_SO}). We assume
that the potential $V({\bf r})$ which is responsible for the SO coupling
consists of two contributions: (i) a parabolic confining potential with a
gradient (or the accompanying electric field) which lies in the plane of
the 2DEG; (ii) a potential which arises from the asymmetry of the quantum
well with the corresponding electric field (Rashba field) being uniform and
directed perpendicular to the device plane. We neglect the crystal-field
(Dresselhaus) contribution to $V({\bf r})$.

The goal of this paper is to reveal the qualitative hallmarks of the SO
interaction in Q1DES rather than to construct a complete and realistic
spectral and transport theory. Therefore we use two major simplifications.
First, we neglect the Coulomb interaction between electrons. At first sight
one may not expect this approximation to work in low-concentration 2D
electron systems where the Coulomb energy may exceed the kinetic energy by
an order of magnitude. However, it has recently been shown
theoretically~\cite{Raikh} that the effect of electron-electron
interactions on SO coupling in such systems can be plausibly taken
into account via a renormalisation of the SO coupling constant. More
specifically, this renormalisation leads to an enhancement (by $10$ --
$50$ \%) of the strength of the SO interaction, which emphasises the
significance of the SO-related effects in low-dimensional electron systems.
Our second simplification is the exploitation of one-band effective mass
approximation~\cite{Ridley} for the Schr\"{o}dinger equation. Within this
approximation, the influence of the crystal forces on electron dynamics in
the conduction band is reduced to the renormalisation of the electron mass
and all interband transitions are left out. Despite the obvious simplicity,
this approach works well~\cite{Kelly,Marques} for a wide range of
semiconductor materials.

Section~\ref{sec-theory} is the central part of the paper. It is devoted to
the solution of the problem that we have outlined above and to the analysis
of the results obtained. In subsection~\ref{subsec-2A} we introduce the
Hamiltonian of a 2DEG that includes the parabolic confining
potential and the SO-interaction term (\ref{H_SO}). We take into account
only those contributions to the SO Hamiltonian that arise from macroscopic
(i.e. relatively smooth) potentials, viz from the quantum-well (Rashba) and
the parabolic confining potentials. Since the operator (\ref{H_SO})
contains the Pauli matrices, we seek electron wave functions in the form of
two-component spinors. Within such a representation the Schr\"{o}dinger
equation turns out to be a system of two {\it coupled} differential
equations with respect to the spinor components. In next two subsections we
solve this system for two physically different situations.
Subsection~\ref{subsec-2B} deals with the case of zero Rashba SO coupling,
where the entire SO interaction comes from the parabolic confining
potential. Here the Schr\"{o}dinger equations decouple and the electron
wave functions and energy spectrum are found in an explicitly analytical
form. In subsection~\ref{subsec-2C} we consider a more general situation
where both mechanisms of the SO-coupling are present. In this case we
calculate the electron wave functions and energy spectrum numerically using
the results of subsection~\ref{subsec-2B} as a convenient basis.
Afterwards we analyse the energy spectrum obtained in order to reveal the
basic features of both SO-coupling mechanisms. In
subsection~\ref{subsec-2D} we discuss possible manifestations of the SO
interaction in the dependence of the ballistic conductance of a Q1DES on
the Fermi energy. Section~\ref{sec-con} summarises the results of our
research.

%%%%%%%%%%%%%%%%%%%%%%%%%%%%% SECTION %%%%%%%%%%%%%%%%%%%%%%%%%%%%%%%%%%%
\section{Theory and analysis}
\label{sec-theory}
%%%%%%%%%%%%%%%%%%%%%%%%%%%%%%%%%%%%%%%%%%%%%%%%%%%%%%%%%%%%%%%%%%%%%%%%%

\subsection{Problem statement. Spinor equation}
\label{subsec-2A}

Within one-band effective mass approximation~\cite{Ridley} the Hamiltonian
of a Q1DES can be written as

\begin{equation}
\hat H =
\frac{\hat{\bf p}^2}{2M}+V_{LC}({\bf r})+\hat{H}_{SO},
\label{H}
\end{equation}
where ${\bf r}=\{x,y\}$ is a 2D position vector and $M$ is the effective
electron mass. In line with the Refs.~\cite{Laux,Drexler,Kardynal}, the
{\it lateral confining potential} $V_{LC}({\bf r})$ is approximated by a
parabola

\begin{equation}
V_{LC}({\bf r})=\frac{M\omega^2}{2}x^2.
\label{V_SG}
\end{equation}
The quantity $\omega$ controls the strength (curvature) of the confining
potential. The in-plane electric field ${\bf E}_{LC}({\bf r})$ associated
with $V_{LC}({\bf r})$ is given by
${\bf E}_{LC}({\bf r})=-\nabla V_{LC}({\bf r})=-M\omega^2{\bf x}$.

We assume that the SO interaction Hamiltonian $\hat{H}_{SO}$ (\ref{H_SO})
in Eq.~(\ref{H}) is formed by two contributions:
$\hat{H}_{SO}=\hat{H}_{SO}^\alpha+\hat{H}_{SO}^\beta$. The first one,
$\hat{H}_{SO}^\alpha$, arises from the asymmetry of the quantum well, i.e.
from the Rashba mechanism~\cite{Rashba,Bychkov} of the SO coupling. Since
the interface-induced (Rashba) electric field can reasonably be assumed
uniform and directed along the $z$-axis, the term $\hat{H}_{SO}^\alpha$ can
be described by the following expression:

\begin{equation}
\hat{H}_{SO}^\alpha = \frac{\alpha}{\hbar}
\left(\hat{\bbox{\sigma}}\times\hat{\bf p}\right)_z=
i\alpha\left(\sigma_y\frac{\partial}{\partial x}
-\sigma_x\frac{\partial}{\partial y}\right).
\label{H_a}
\end{equation}
The SO-coupling constant $\alpha$ takes values within a range ($1$ --
$10$)$\times 10^{-10}$ eV$\times$cm for different
systems~\cite{Das,Nitta,Luo,Bychkov,Hassenkam}. For brevity, in what
follows we will refer to the Rashba mechanism of the SO coupling as
$\alpha$-coupling.

The second contribution $\hat{H}_{SO}^\beta$ to $\hat H_{SO}$ comes from
the parabolic confining potential (\ref{V_SG}):

\begin{equation}
\hat{H}_{SO}^\beta = \frac{\beta}{\hbar}\frac{x}{l_\omega}
\left(\hat{\bbox{\sigma}}\times
\hat{\bf p}\right)_x =
i\beta\frac{x}{l_\omega}\sigma_z
\frac{\partial}{\partial y}.
\label{H_b}
\end{equation}
Here $l_\omega=(\hbar/M\omega)^{1/2}$ is the typical spatial scale
associated with the potential $V_{LC}$ (\ref{V_SG}). In Eq.~(\ref{H_b}) we
have introduced the SO-coupling constant $\beta$. Comparison of
typical electric fields originated from the quantum-well and lateral
confining potentials allows one to conclude that a plausible estimate for
$\beta$ should be roughly $10$\% of $\alpha$. Moreover, in square quantum
wells where the value of $\alpha$ is considerably
diminished~\cite{Chen,Hassenkam} (by an order of magnitude) the constant
$\beta$ may well compete with $\alpha$. Henceforth we adopt the term
$\beta$-coupling for the mechanism of the SO-interaction arising from the
lateral confining potential (\ref{V_SG}).

Our objective is to find eigenvalues and eigenfunctions of the
Schr\"{o}dinger equation $\hat H\Psi=E\Psi$ with the Hamiltonian $\hat H$
given by Eqs.~(\ref{H}) -- (\ref{H_b}). The wave function
$\Psi=\Psi({\bf r})=\{\Psi_\uparrow({\bf r})\;\Psi_\downarrow({\bf r})\}$
is a two-component spinor and the energy $E$ is measured from the conduction
band edge. It is easy to see that the Hamiltonian (\ref{H}) -- (\ref{H_b})
is translationally invariant in the $y$-direction. We therefore seek
solutions to Schr\"{o}dinger's equation in the form of plane waves
propagating along the $y$-axis, i.e. $\Psi_{\uparrow\downarrow} ({\bf
r})=\exp(ik_yy)\Phi_{\uparrow\downarrow}(t)$. Here $t=x/l_\omega$ is the
dimensionless transverse coordinate, $\Phi_{\uparrow\downarrow}(t)$ is the
transverse wave function, and $k_y$ is the longitudinal wave number. After
substituting this representation into the Schr\"{o}dinger equation, we
obtain a system of two differential equations with respect to the spinor
components $\Phi_{\uparrow\downarrow}(t)$:

\begin{eqnarray}
&\Phi_\uparrow^{\prime\prime}+
\left(\varepsilon_x-t^2+t_\beta\, t\right)
\Phi_\uparrow(t)=
\nonumber \\
&(l_\omega/l_\alpha)
\left[(k_yl_\omega)\Phi_\downarrow(t) + \Phi_\downarrow^\prime \right]
\label{plus} \\
\nonumber \\
&\Phi_\downarrow^{\prime\prime}+
\left(\varepsilon_x-t^2-t_\beta\, t\right)
\Phi_\downarrow(t)=
\nonumber \\
&(l_\omega/l_\alpha)
\left[(k_yl_\omega)\Phi_\uparrow(t) - \Phi_\uparrow^\prime \right],
\label{minus}
\end{eqnarray}

\begin{equation}
t_\beta=\frac{l_\omega}{l_\beta}(k_yl_\omega),
\label{tb}
\end{equation}
where $\varepsilon_x \equiv (k_xl_\omega)^2$ is the dimensionless
transverse energy, $k_x^2=(2M/\hbar^2)E-k_y^2$, and the prime denotes a
derivative with respect to $t$. The lengths $l_\alpha$ and $l_\beta$
defined by

\begin{equation}
l_\alpha=\hbar^2/2M\alpha, \qquad
l_\beta=\hbar^2/2M\beta
\label{l_ab}
\end{equation}
are characteristic spatial scales associated with the $\alpha$- and
$\beta$-couplings, respectively. We note that the functions
$\Phi_{\uparrow\downarrow}(t)$ depend on three dimensionless external
parameters: $l_\omega/l_\alpha$, $l_\omega/l_\beta$, and $k_yl_\omega$.

The equations (\ref{plus}) and (\ref{minus}) are arranged in such a way that
all the terms which couple them together are collected on the rhs'.
It is interesting that this arrangement separates explicitly the $\alpha$
and $\beta$ mechanisms of the SO interaction. Indeed, the $\alpha$-terms
enter only the rhs', while all the $\beta$-terms are contained on the lhs'.
This suggests that the $\beta$-coupling is responsible for forming
independent (``unperturbed'' or ``non-interacting'') wave functions
$\Phi_{\uparrow\downarrow}(t)$, while the $\alpha$-coupling mixes them
together to form the solution of the whole system (\ref{plus}) and
(\ref{minus}).

\subsection{Zero $\alpha$-coupling (square quantum well)}
\label{subsec-2B}

We start the analysis of Eqs.~(\ref{plus}) and (\ref{minus}) with the
situation where the quantum well is square so that the interface-induced
electric field vanishes. In terms of the SO-interaction this means that the
Rashba mechanism (i.e. the $\alpha$-mechanism) of the SO-coupling can be
omitted. In other words, the characteristic length $l_\alpha$ of the
$\alpha$-interaction tends to infinity: $l_\alpha \to \infty$, or
$l_\omega/l_\alpha \to 0$. The study of this relatively simple case will
provide us with a convenient basis for treating the general problem with
finite $l_\alpha$.

Once $l_\omega/l_\alpha$ has been put equal to zero, the rhs' in
Eqs.~(\ref{plus}) and (\ref{minus}) vanish and they transform into two
independent Hermite equations~\cite{Abramowitz} whose exact solutions are

\begin{eqnarray}
\phi_{\uparrow\downarrow}^n(t)&\equiv&
\left.\Phi_{\uparrow\downarrow}^n(t)\right|_{l_\omega/l_\alpha=0}
\nonumber \\
&=&
\frac{\exp\left[-(t \mp t_\beta/2)^2/2\right]}
{\sqrt{2^n n!\pi^{1/2}}}
H_n\left(t \mp t_\beta\right) \nonumber \\
&& \qquad \qquad \qquad \qquad  n=0,1,2,\ldots
\label{phi0}
\end{eqnarray}
Here $H_n(t)$ is the Hermite polynomial of $n$-th order. The real wave
functions $\phi_\uparrow^n(t)$ and $\phi_\downarrow^n(t)$ form complete sets
with respect to the discrete quantum number $n$ and are normalised by the
following conditions:

\begin{equation}
\left<\phi_{\uparrow\downarrow}^m |
\phi_{\uparrow\downarrow}^n\right>\equiv
\int\limits_{-\infty}^\infty dt\,
\phi_{\uparrow\downarrow}^m(t)\phi_{\uparrow\downarrow}^n(t)=
\delta_{mn},
\label{norm}
\end{equation}
where both $\phi_{\uparrow\downarrow}^m$ and $\phi_{\uparrow\downarrow}^n$
are taken at a same value of $k_y$.

\vspace{0.5cm}
\begin{figure}
\begin{center}
\epsfig{file=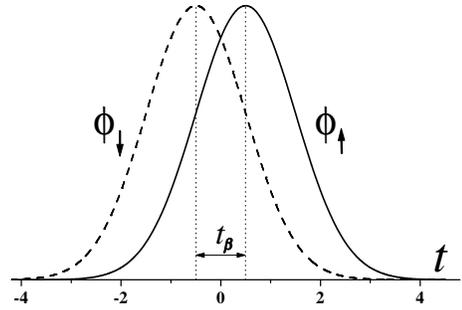,width=2.50in}
\caption{The spinor components $\phi_\uparrow(t)$ and $\phi_\downarrow(t)$
for $n=0$, $l_\omega/l_\beta=0.1$, $l_\omega/l_\alpha=0.0$,
$k_yl_\omega=10.0$.}
\label{fig1}
\end{center}
\end{figure}

From Eq.~(\ref{phi0}) one can see that the both ``up'' ($\phi_\uparrow^n$)
and ``down'' ($\phi_\downarrow^n$) spinor components have exactly the same
shape as functions of $t$ but are spatially displaced with respect to each
other by an amount of $t_\beta$ (see Fig.~\ref{fig1}). According to the
definition (\ref{tb}), this displacement is a direct consequence of the
$\beta$-coupling because it does not vanish as long as
$l_\omega/l_\beta \neq 0$ (except where $k_yl_\omega=0$). At finite values
of $t_\beta$, the complete sets of functions $\phi_\uparrow^n(t)$ and
$\phi_\downarrow^n(t)$ turn out {\it not} to be mutually orthogonal, i.e.
 $\left<\phi_{\uparrow\downarrow}^m |
\phi_{\downarrow\uparrow}^n\right>\neq \delta_{mn}$. Instead, the scalar
cross-product $\left<\phi_{\uparrow\downarrow}^m |
\phi_{\downarrow\uparrow}^n\right>$ in the case of sufficiently weak
$\beta$-coupling ($t_\beta \ll 1$) is governed by the following
asymptotics:

\begin{equation}
\left<\phi_{\uparrow\downarrow}^n |
\phi_{\downarrow\uparrow}^n\right> = 1 - O(t_\beta^2),
\label{nn}
\end{equation}

\begin{equation}
\left<\phi_{\uparrow\downarrow}^n |
\phi_{\downarrow\uparrow}^{n+1}\right> \simeq
\pm\sqrt{\frac{n+1}{2}}\;t_\beta =
\pm\sqrt{\frac{n+1}{2}}\;\frac{l_\omega}{l_\beta}(k_yl_\omega),
\label{nn+1}
\end{equation}

\begin{equation}
\left<\phi_{\uparrow\downarrow}^n |
\phi_{\downarrow\uparrow}^{n-1}\right> \simeq
\mp\sqrt{\frac{n}{2}}\;t_\beta =
\mp\sqrt{\frac{n}{2}}\;\frac{l_\omega}{l_\beta}(k_yl_\omega),
\label{nn-1}
\end{equation}

\begin{equation}
\left<\phi_{\uparrow\downarrow}^n |
\phi_{\downarrow\uparrow}^{n \pm p}\right>
= O(t_\beta^p)
\quad\mbox{for}\quad
p \geq 2.
\label{nn+p}
\end{equation}
The ``displacing effect'' of the $\beta$-coupling on spinor wave functions
[see Eq.~\ref{phi0} and Fig.~\ref{fig1}] is superficially similar to the
effect of a perpendicular magnetic field on a Q1DES. However, the essential
difference is that a magnetic field shifts {\it both} spinor components
$\phi_{\uparrow}^n(t)$ and $\phi_{\downarrow}^n(t)$ {\it as a whole} by an
amount proportional to the strength of the magnetic field (see, e.g.,
Ref.~\cite{Landau}), but it does {\it not} affect their {\it mutual}
spatial distribution. In contrast to this, as can be seen in
Fig.~\ref{fig1}, the $\beta$-coupling causes a spatial separation of the
spin-polarised electron states $\phi_{\uparrow}^n(t)$ and
$\phi_{\downarrow}^n(t)$.

\begin{figure}
\begin{center}
\epsfig{file=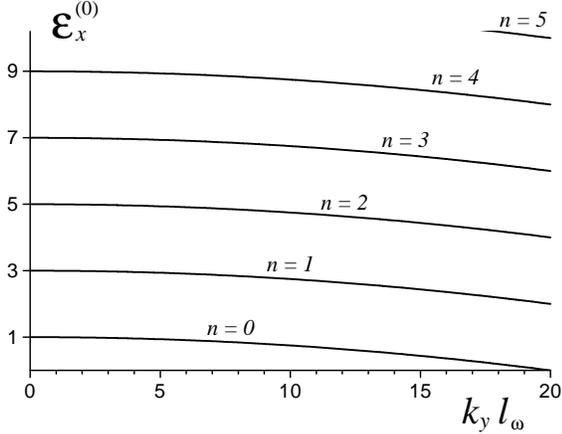,width=3.00in}
\caption{The transverse energy $\varepsilon_x^{(0)}$ vs. $k_yl_\omega$ for
$l_\omega/l_\beta=0.1$.}
\label{fig2}
\end{center}
\end{figure}

Although the $\beta$-coupling shifts the spinor components
$\phi_{\uparrow\downarrow}^n(t)$ apart, it nevertheless is not capable of
lifting their energy degeneracy. The dimensionless transverse energy
$\varepsilon_x^{(0)}$ corresponding to the both eigenfunctions
$\phi_{\uparrow\downarrow}^n(t)$ (\ref{phi0}) is given by (see
Fig.~\ref{fig2})

\begin{eqnarray}
\varepsilon_x^{(0)}\equiv\varepsilon_x^{(0)}(n,k_y)&=&
2n+1-\left(t_\beta/2\right)^2  \nonumber \\
&=&
2n+1-\frac{1}{4}\left(\frac{l_\omega}{l_\beta}\right)^2(k_yl_\omega)^2.
\label{spec0}
\end{eqnarray}
The total electron energy $E$ then forms parabolic subbands for each $n$-th
transverse mode:

\begin{equation}
E \equiv E_n(k_y) =
\frac{\hbar\omega}{2}\varepsilon_x^{(0)}(n,k_y) +
\frac{\hbar^2k_y^2}{2M}.
\label{E}
\end{equation}
For zero $\beta$-coupling ($l_\omega/l_\beta=0$) formulas (\ref{spec0}) and
(\ref{E}) describe the well-known electric subbands~\cite{Datta,Thankappan}.

\subsection{Finite $\alpha$-coupling (triangular quantum well)}
\label{subsec-2C}

We now examine the situation where the effective geometry of the quantum
well is such that the interface-induced electric field is non-zero. An
example is the triangular well which forms at a heterojunction
interface~\cite{Kelly,Ando}. This interface-induced field gives rise to a
finite $\alpha$-coupling and therefore to finite values of the coupling
constant $l_\omega/l_\alpha$. For arbitrary non-zero values of
$l_\omega/l_\alpha$, eigenvalues of Eqs.~(\ref{plus}) and (\ref{minus})
cannot be obtained as simply as they were in the case of
$l_\omega/l_\alpha=0$. To analyse the energy spectrum in the presence of
$\alpha$-coupling, we will follow the scheme proposed in Ref.~\cite{BMR}.
Namely, we first expand the unknown functions
$\Phi_{\uparrow\downarrow}(t)$ in terms of the unperturbed solutions
$\phi_{\uparrow\downarrow}(t)$ (\ref{phi0}) of Eqs.~(\ref{plus}) and
(\ref{minus}):

\begin{equation}
\Phi_{\uparrow\downarrow}(t)=\sum_{m=0}^\infty
f_{\uparrow\downarrow}^m\;\phi_\uparrow^m(t),
\label{exp1}
\end{equation}

\begin{equation}
\Phi_{\uparrow\downarrow}(t)=\sum_{m=0}^\infty
g_{\uparrow\downarrow}^m\;\phi_\downarrow^m(t).
\label{exp2}
\end{equation}
We then substitute the expansions (\ref{exp1}) and (\ref{exp2}) into
Eqs.~(\ref{plus}) and (\ref{minus}) respectively and make use of the
property~\cite{Abramowitz}

\begin{equation}
\frac{d}{dt}\phi_{\uparrow\downarrow}^n(t)=
\sqrt{\frac{n}{2}}\phi_{\uparrow\downarrow}^{n-1}(t)-
\sqrt{\frac{n+1}{2}}\phi_{\uparrow\downarrow}^{n+1}(t)
\label{deriv-prop}
\end{equation}
to remove the derivative with respect to $t$ on the rhs' of
Eqs.~(\ref{plus}) and (\ref{minus}). We next multiply the equations
obtained by $\phi_\uparrow^n(t)$ and $\phi_\downarrow^n(t)$ respectively
and integrate them over $t$ from $-\infty$ to $\infty$. Owing to the
orthogonality condition (\ref{norm}), the summation over $m$ is removed by
the delta-function $\delta_{mn}$ and we find the simple vector relations:

\begin{equation}
{\bf f}_\uparrow = \hat{U}_\uparrow \; {\bf f}_\downarrow,
\label{rel1}
\end{equation}

\begin{equation}
{\bf g}_\downarrow = \hat{U}_\downarrow \; {\bf g}_\uparrow.
\label{rel2}
\end{equation}
Here ${\bf f}_{\uparrow\downarrow} \equiv
\{f_{\uparrow\downarrow}^0,f_{\uparrow\downarrow}^1,\ldots,
f_{\uparrow\downarrow}^n,\ldots\}$ and ${\bf g}_{\uparrow\downarrow} \equiv
\{g_{\uparrow\downarrow}^0,g_{\uparrow\downarrow}^1,\ldots,
g_{\uparrow\downarrow}^n,\ldots\}$ are vectorised coefficients of the
expansions (\ref{exp1}) and (\ref{exp2}) respectively, and
$\hat{U}_{\uparrow\downarrow}$ are tridiagonal matrices defined by their
elements:

\begin{equation}
U_{\uparrow\downarrow}^{nn}=\frac{l_\omega}{l_\alpha}\;
\frac{k_yl_\omega}{\varepsilon_x-\varepsilon_x^{(0)}(n)},
\label{U-diag}
\end{equation}

\begin{equation}
U_{\uparrow\downarrow}^{n,n+1}=\pm\frac{l_\omega}{l_\alpha}
\left(\frac{n+1}{2}\right)^{1/2}\frac{1}
{\varepsilon_x-\varepsilon_x^{(0)}(n)},
\label{U-upper}
\end{equation}

\begin{equation}
U_{\uparrow\downarrow}^{n+1,n}=\mp\frac{l_\omega}{l_\alpha}
\left(\frac{n+1}{2}\right)^{1/2}\frac{1}
{\varepsilon_x-\varepsilon_x^{(0)}(n+1)}.
\label{U-lower}
\end{equation}
Note that the matrices $\hat{U}_{\uparrow\downarrow}$ are neither symmetric
nor anti-symmetric.

Eqs.~(\ref{rel1}) -- (\ref{U-lower}) illustrate the effectiveness of the
representations (\ref{exp1}) and (\ref{exp2}). Indeed, by using the
functions $\phi_{\uparrow\downarrow}^m(t)$ as expansion bases, we have
reduced differential operators on the lhs' of Eqs.~(\ref{plus}) and
(\ref{minus}) to the scalar factor $\varepsilon_x-\varepsilon_x^{(0)}(n)$.
In other words, matrices which were supposed to act on ${\bf f}_\uparrow$
and ${\bf g}_\downarrow$ in Eqs.~(\ref{rel1}) and (\ref{rel2}) turn out to
be diagonal within the representations (\ref{exp1}) and (\ref{exp2}).

Our next step is to find the relationship between the vectors
${\bf f}_{\uparrow\downarrow}$ and ${\bf g}_{\uparrow\downarrow}$. To do
this, we equate the rhs of Eq.~(\ref{exp1}) to that of Eq.~(\ref{exp2}) and
take the scalar product of the resulting equation with
$\phi_\uparrow^n(t)$.  After making use of the orthogonalisation condition
(\ref{norm}), we find that

\begin{equation}
{\bf f}_{\uparrow\downarrow} = \hat{W}_\uparrow \;
{\bf g}_{\uparrow\downarrow},
\label{rel3}
\end{equation}
where the matrix $\hat{W}_\uparrow$ is defined by

\begin{equation}
W_{\uparrow\downarrow}^{mn}=
\left<\phi_{\uparrow\downarrow}^m |
\phi_{\downarrow\uparrow}^n\right>.
\label{W}
\end{equation}
According to Eq.~(\ref{phi0}) [see also Eqs.~(\ref{nn}) -- (\ref{nn+p})]
the matrices $\hat W_{\uparrow\downarrow}$ are {\it not} diagonal as long
as the $\beta$-coupling is {\it finite}.

Analogously, by taking a scalar product of the rhs' of Eqs.~(\ref{exp1}) and
(\ref{exp2}) by $\phi_\downarrow^n(t)$, it can be shown that

\begin{equation}
{\bf g}_{\uparrow\downarrow} = \hat{W}_\downarrow \;
{\bf f}_{\uparrow\downarrow}.
\label{rel4}
\end{equation}

Finally, we combine relations (\ref{rel1}), (\ref{rel2}), (\ref{rel3}), and
(\ref{rel4}) into a closed homogeneous equation with respect to
${\bf f}_\uparrow$:
${\bf f}_\uparrow =
\hat{U}_\uparrow\hat{W}_\uparrow\hat{U}_\downarrow\hat{W}_\downarrow \;
{\bf f}_\uparrow$. In order for this equation to have a non-trivial
solution, the Jacobian matrix must satisfy the following condition:

\begin{equation}
{\rm det}(1-\hat{U}_\uparrow\hat{W}_\uparrow
\hat{U}_\downarrow\hat{W}_\downarrow)=0.
\label{det}
\end{equation}
The roots $\varepsilon_x$ of this equation determine the dispersion law
of electrons (see, e.g., Ref.~\cite{BMR}). Here,
$\varepsilon_x \to \varepsilon_x^{(0)}$ as $l_\omega/l_\alpha \to 0$.

Alternatively, the relations (\ref{rel1}), (\ref{rel2}), (\ref{rel3}), and
(\ref{rel4}) could have been resolved with respect to ${\bf g}_\downarrow$,
which would have led to the permuted Jacobian matrix
$\hat{U}_\downarrow\hat{W}_\downarrow\hat{U}_\uparrow\hat{W}_\uparrow$.
However, such a permutation leaves the determinant unchanged and
therefore the dispersion equation (\ref{det}) would have been still
applicable.

To obtain the solution of Eq.~(\ref{det}) for given values of the external
parameters $l_\omega/l_\alpha$, $l_\omega/l_\beta$, and $k_yl_\omega$ we
truncated the infinite matrix $1-\hat{U}_\uparrow\hat{W}_\uparrow
\hat{U}_\downarrow\hat{W}_\downarrow$ down to the first $N$ rows and
columns and used a numerical root finder to obtain the zeros of its
determinant. Owing to the conveniently chosen
bases (\ref{exp1}) and (\ref{exp2}), the roots of the equation (\ref{det})
converge very quickly to their exact values as $N$ is increased. For
instance, for $l_\omega/l_\alpha$, $l_\omega/l_\beta \lesssim 1$, it
suffices to take $N \approx 30$ to find the ten lowest energy levels with
a net error $<10^{-8}$.

The energy spectrum $\varepsilon_x=\varepsilon_x(k_yl_\omega)$ calculated
by this procedure is shown in Fig.~\ref{fig3}. Let us
first consider the case of negligibly weak $\beta$-coupling, when
$l_\omega/l_\beta \to 0$ [Fig.~\ref{fig3}(a)]. Here we see two-fold spin
degeneracy of all quantum levels at $k_y=0$. As soon as we move away
from the point $k_y=0$, the SO interaction lifts the degeneracy and
produces an energy splitting
$\Delta_R=\varepsilon_x^\uparrow-\varepsilon_x^\downarrow$ proportional to
$k_y$, where $\varepsilon_x^{\uparrow\downarrow}$ are energies associated
with the ``up'' and ``down'' spin polarisations respectively. This linear
behaviour agrees with both theoretical
predictions~\cite{Raikh,Rashba,Bychkov,BMR} and experimental
observations~\cite{Das,Chen,Nitta,Luo} of the Rashba splitting in $2D$
systems.

\vspace{0.5cm}
\begin{figure}
\begin{center}
\epsfig{file=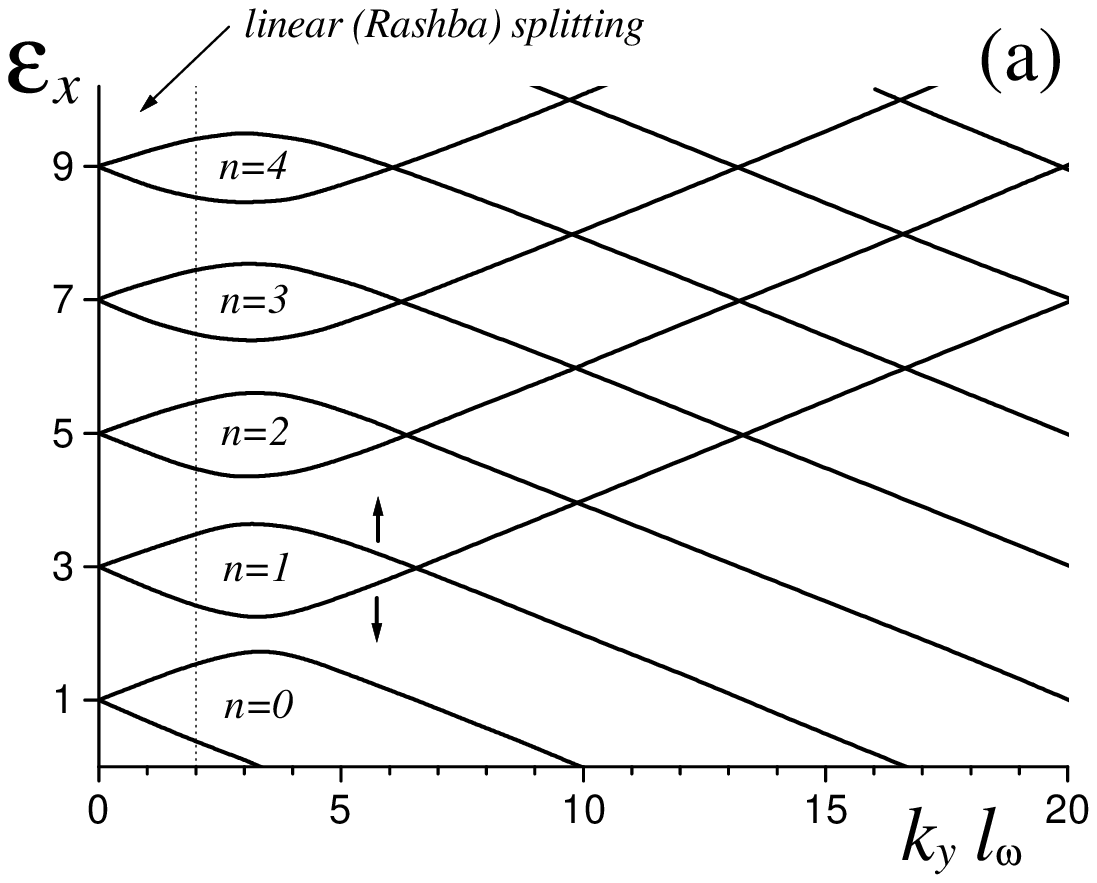,width=3.00in}
\end{center}
\end{figure}

\begin{figure}
\begin{center}
\epsfig{file=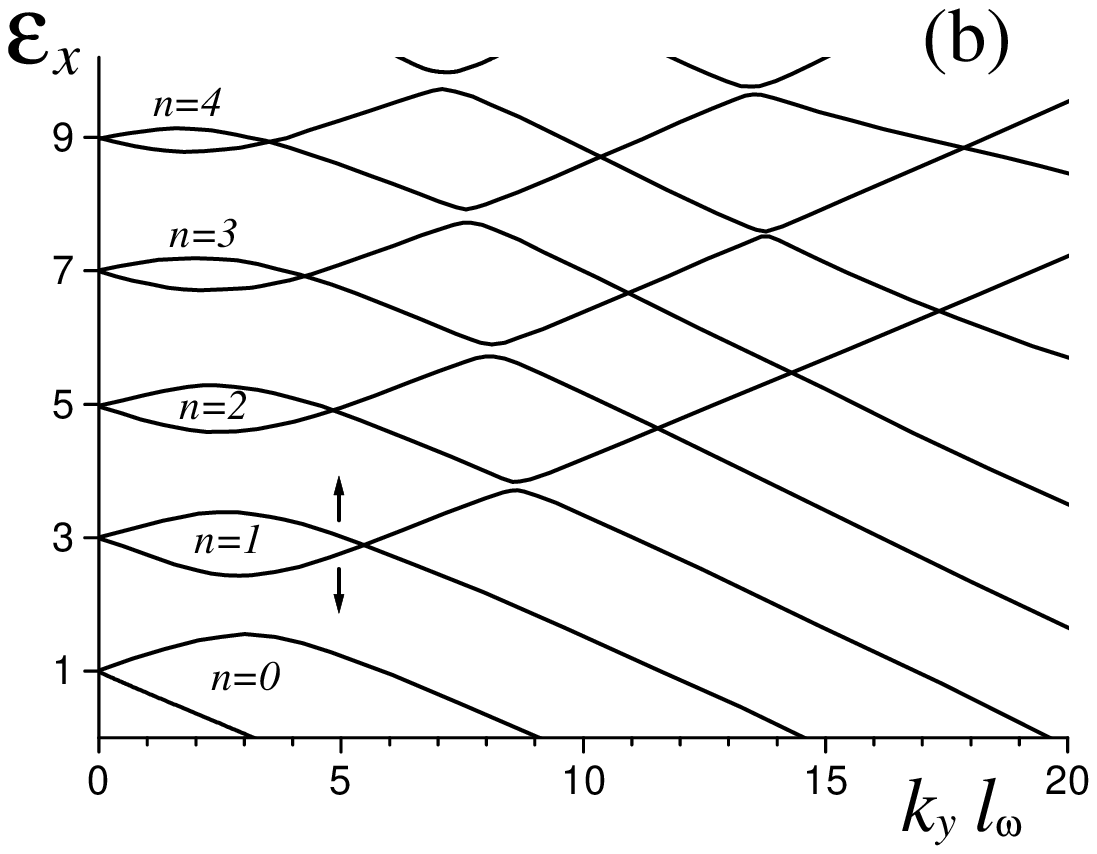,width=3.00in}
\caption{The transverse energy $\varepsilon_x$ vs. $k_yl_\omega$ for finite
$\alpha$-coupling ($l_\omega/l_\alpha=0.3$): (a) $l_\omega/l_\beta=0$; (b)
$l_\omega/l_\beta=0.1$.}
\label{fig3}
\end{center}
\end{figure}

The linear dependence of $\Delta_R$ on $k_y$ can be easily deduced
analytically from Eqs.~(\ref{plus}) and (\ref{minus}) within first-order
perturbation theory with respect to the parameter $l_\omega/l_\alpha$.
Indeed, if we treat the rhs' of Eqs.~(\ref{plus}) and (\ref{minus}) as
perturbations, then the first-order correction
$\Delta\varepsilon_x^{\uparrow\downarrow}=
\varepsilon_x^{\uparrow\downarrow}-\varepsilon_x^{(0)}$ to the
unperturbed energy (\ref{spec0}) will be given by

\begin{equation}
\Delta\varepsilon_x^{\uparrow\downarrow} \simeq
\pm\sqrt{\hat{P}_\uparrow^{nn}\hat{P}_\downarrow^{nn}},
\label{Deps}
\end{equation}
or

\begin{equation}
\Delta_R=\varepsilon_x^\uparrow-\varepsilon_x^\downarrow \simeq
2\sqrt{P_\uparrow^{nn}P_\downarrow^{nn}}.
\label{DR0}
\end{equation}
Here $P_{\uparrow\downarrow}^{nn}$ are diagonal matrix elements of
perturbation operators $\hat{P}_{\uparrow\downarrow}$:

\begin{equation}
P_{\uparrow\downarrow}^{mn} \equiv
\left<\phi_{\uparrow\downarrow}^m |
\hat{P}_{\uparrow\downarrow} |
\phi_{\downarrow\uparrow}^n\right>,
\label{Pmn}
\end{equation}

\begin{equation}
\hat{P}_{\uparrow\downarrow} =\frac{l_\omega}{l_\alpha}\left(
\pm\frac{d}{dt}+k_yl_\omega\right).
\label{P0}
\end{equation}
To calculate the matrix elements $P_{\uparrow\downarrow}^{nn}$ we use the
property (\ref{deriv-prop}) and take into account that

\begin{equation}
\left<\phi_{\uparrow\downarrow}^m |
\phi_{\downarrow\uparrow}^n\right>=\delta_{mn}
\qquad\mbox{for}\qquad
l_\omega/l_\beta=0
\label{norm2}
\end{equation}
[see Eqs.~(\ref{nn}) -- (\ref{nn+p})]. As a result, we obtain
$P_{\uparrow\downarrow}^{nn} \simeq (l_\omega/l_\alpha)(k_yl_\omega)$ and
hence the Rashba splitting $\Delta_R$ (\ref{DR0}) is governed by

\begin{equation}
\Delta_R \simeq
2\frac{l_\omega}{l_\alpha}(k_yl_\omega)
\qquad\mbox{for}\qquad
l_\omega/l_\beta=0,
\label{DR1}
\end{equation}
i.e. it is indeed proportional to the longitudinal wave number $k_y$.

The perturbative result (\ref{DR1}) applies as long as
$\Delta_R \ll \varepsilon_x^{(0)}(n+1)-\varepsilon_x^{(0)}(n)$, i.e. if
$(l_\omega/l_\alpha)(k_yl_\omega) \ll 1$. For $l_\omega/l_\alpha=0.3$ this
condition restricts $k_yl_\omega$ to values much less than about $3$. In
Fig.~\ref{fig3}(a) we see that the linear behaviour of the energy splitting
for all the curves holds true until $k_yl_\omega \approx 2$ (dotted line).
Within the region $k_yl_\omega \lesssim 2$ the asymptotic (\ref{DR1}) gives
a very good fit to all the energy branches. However, as soon as
$k_yl_\omega$ becomes larger than $\approx 2.5$, the dispersion curves
start to bend. Eventually this leads to an {\it anticrossing} of branches
corresponding to quantum levels with neighbouring discrete numbers $n$.
This fact drastically contrasts the simpler situation with no confining
potential (\ref{V_SG}), where the linearly split spectrum (\ref{DR1})
represents the exact solution~\cite{Bychkov}.

Usually~\cite{Landau} anticrossing is associated with similar symmetry
(``hybridisation'') of underlying states (wave functions). To reveal
possible symmetries in our case, we return to the perturbation theory
with respect to $l_\omega/l_\alpha$ and consider first-order corrections
$\delta\Phi_{\uparrow\downarrow}^n(t)$ to the wave functions (\ref{phi0}):

\begin{equation}
\delta\Phi_{\uparrow\downarrow}^n(t) = \sum_{m \neq n}
\frac{P_{\uparrow\downarrow}^{mn}}
{\varepsilon_x^{(0)}(n)-\varepsilon_x^{(0)}(m)}
\phi_{\uparrow\downarrow}^m(t).
\label{dphi}
\end{equation}

Owing to the orthogonality condition (\ref{norm2}) the second (constant)
term in Eq.~(\ref{P0}) does {\it not} contribute to
$\delta\Phi_{\uparrow\downarrow}^n(t)$ (\ref{dphi}) for any $m \neq n$. At
the same time, according to Eqs.~(\ref{deriv-prop}) and (\ref{norm2}), the
first (differential) term in Eq.~(\ref{P0}) gives rise to {\it non-zero}
matrix elements $P_{mn}$ for $m=n+1$ and $m=n-1$. As a result,
$\delta\Phi_{\uparrow\downarrow}^n(t)$ (\ref{dphi}) takes the form:

\begin{eqnarray}
&\delta\Phi_{\uparrow\downarrow}^n(t) \simeq
\pm\frac{1}{2\sqrt{2}}\frac{l_\omega}{l_\alpha}
\left\{\sqrt{n}\;\phi_{\uparrow\downarrow}^{n-1}(t)+
\sqrt{n+1}\;\phi_{\uparrow\downarrow}^{n+1}(t)\right\}
\nonumber \\
&\mbox{for}\qquad l_\omega/l_\beta \to 0.
\label{dphi0}
\end{eqnarray}
The total (perturbed) wave function is given by
$\Phi_{\uparrow\downarrow}^n=\phi_{\uparrow\downarrow}^n+
\delta\Phi_{\uparrow\downarrow}^n$. Fig.~\ref{fig4} shows graphs of
$\Phi_{\uparrow\downarrow}^n(t)$ for $n=0$ and $n=1$.

\vspace{0.5cm}
\begin{figure}
\begin{center}
\epsfig{file=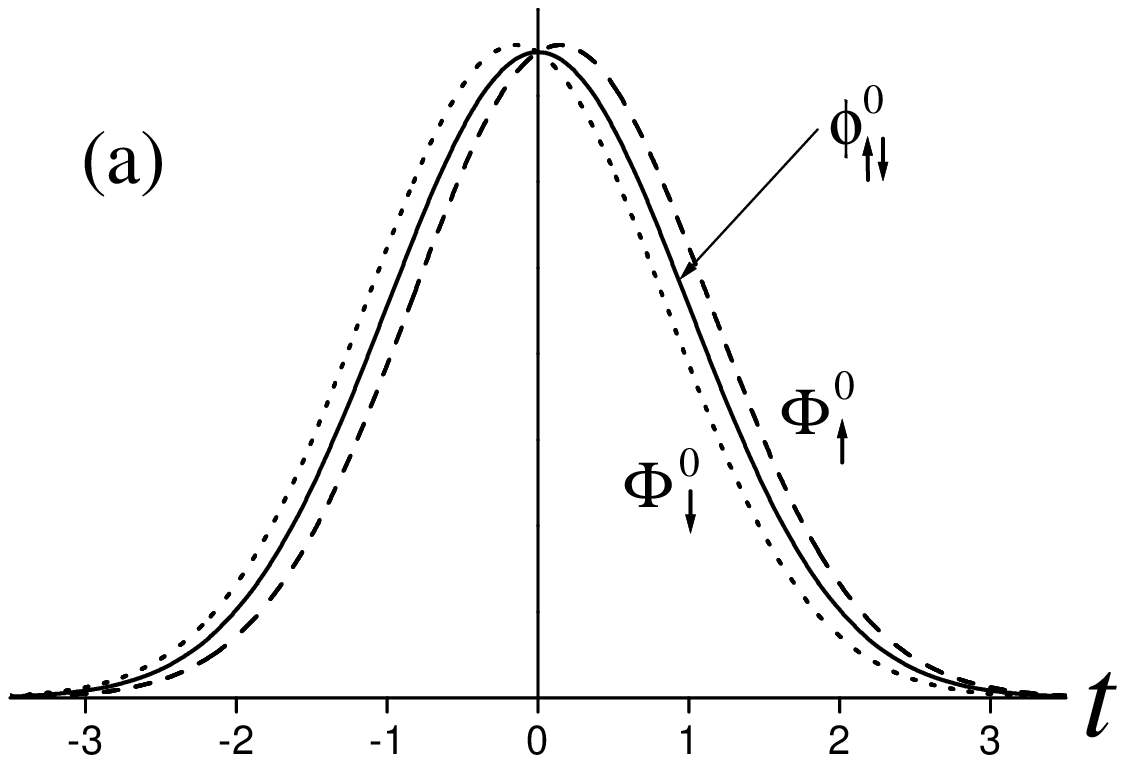,width=3.00in}
\end{center}
\end{figure}

\begin{figure}
\begin{center}
\epsfig{file=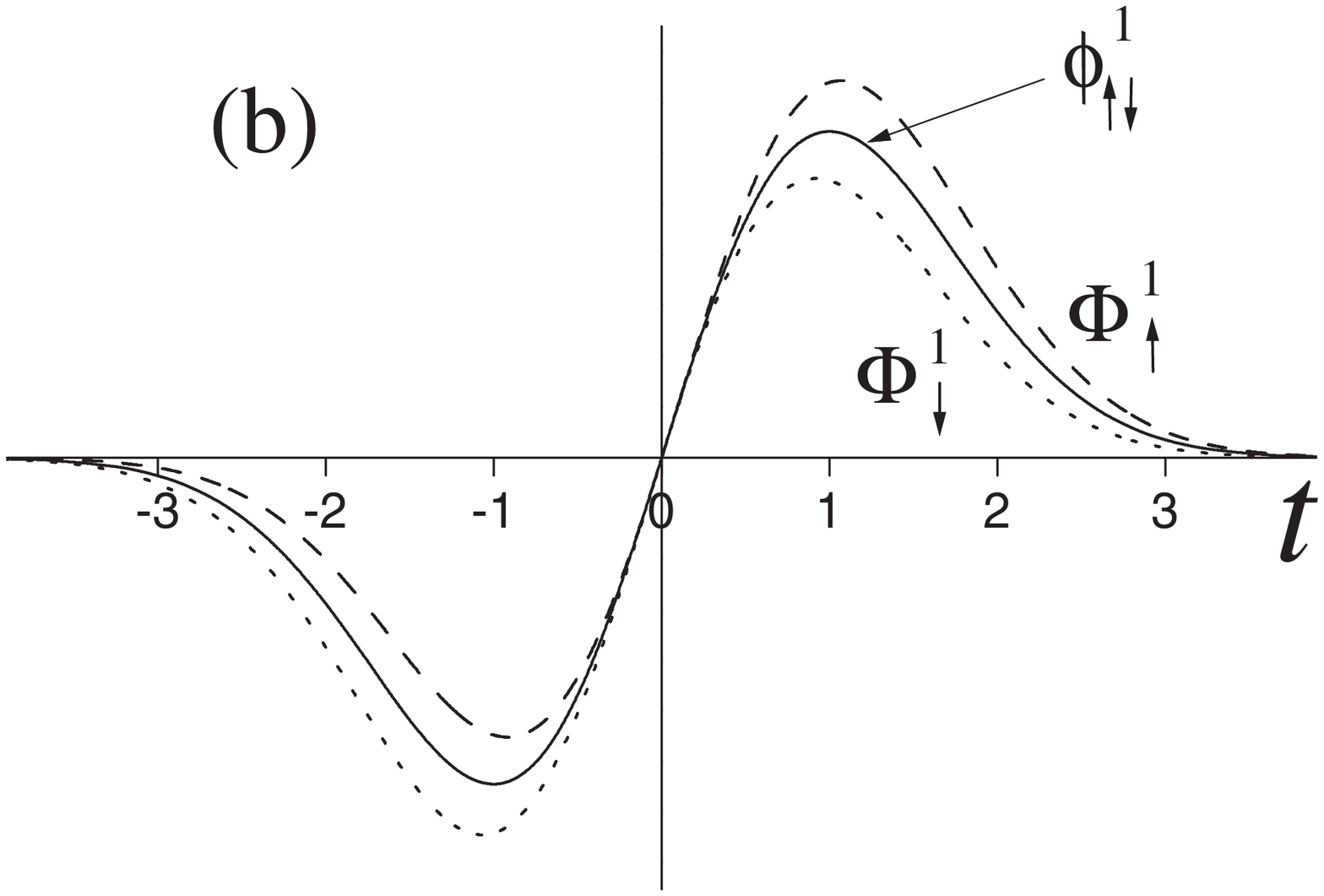,width=3.00in}
\caption{Unperturbed wave functions ($l_\omega/l_\alpha=0$, solid curve) and
wave functions modified by the $\alpha$-coupling ($l_\omega/l_\alpha=0.3$,
the dashed curve describes the ``up'' and the dotted curve the ``down''
spin orientations): (a) $n=0$; (b) n=1. Zero $\beta$-coupling
($l_\omega/l_\beta=0$) is assumed. For definiteness, we take
$k_yl_\omega=0$.}
\label{fig4}
\end{center}
\end{figure}

Eq.~(\ref{dphi0}) demonstrates that the $n$-th quantum state is no longer
independent of its ``neighbours'', i.e. of the $(n \pm 1)$-st states. In
turn, the $(n+1)$-st state now depends on its own nearest neighbours, i.e.
on both the $n$-th and the $(n+2)$-nd states. Owing to this interstate
coupling, all spinors $\Phi^n$ partially acquire the symmetry properties of
spinors $\Phi^{n \pm 1}$. Moreover, since the result (\ref{dphi0}) is
perturbative, it is plausible that the interstate coupling becomes more
pronounced when the perturbative approach breaks down, i.e. when
$k_yl_\omega \gtrsim 2.5$. This ``hybridisation'' of electron states
accounts for the anticrossing of neighbouring energy branches in
Fig.~\ref{fig3}(a).

After passing the relatively narrow anticrossing region $2.5 \lesssim
k_yl_\omega \lesssim 4.5$, all the curves in Fig.~\ref{fig3}(a) straighten
out and adopt a very accurate linear behaviour for $k_yl_\omega \gtrsim 5$.
There are two main features that attract attention in the infinite interval
$k_yl_\omega \gtrsim 5$: (i) there are no further anticrossings of
approaching energy branches. Even those with closest quantum numbers $n$,
which anticrossed at smaller values of $k_yl_\omega$, now cross; (ii) all
the straight lines $\varepsilon_x^{\uparrow\downarrow}(n)$ go down (up)
with the same slope $\approx \mp l_\omega/l_\alpha$ which is independent of
a level number $n$. Although statement (i) means asymptotical smallness
(rather than absolute absence) of anticrossings in the region of large
$k_yl_\omega \gtrsim 5$ in Fig.~\ref{fig3}(a), these anticrossings are much
weaker than any other anticrossings in the system and therefore can be left
out.

To explain these properties we should return to Eqs.~(\ref{plus}),
(\ref{minus}), and (\ref{phi0}). According to Eq.~(\ref{phi0}), the spatial
variation scale of functions $\phi_{\uparrow\downarrow}^n(t)$ is of the
order unity. This means that the derivatives
$\Phi_{\downarrow\uparrow}^\prime$ on the rhs' of Eqs.~(\ref{plus}) and
(\ref{minus}) respectively can be roughly estimated as
$\Phi_{\downarrow\uparrow}$, at least for $|t| \lesssim 1$ and sufficiently
small values of $l_\omega/l_\alpha$. Hence, it seems plausible that within
the region, where $k_yl_\omega \gg 1$, the terms
$\Phi_{\downarrow\uparrow}^\prime$ can be neglected in comparison with
$(k_yl_\omega)\Phi_{\downarrow\uparrow}$. Taking this into account together
with the condition $l_\omega/l_\beta=0$, we reduce Eqs.~(\ref{plus}) and
(\ref{minus}) to

\begin{equation}
\Phi_\uparrow^{\prime\prime}+
\left(\varepsilon_x-t^2\right)\Phi_\uparrow(t)=
(l_\omega/l_\alpha)(k_yl_\omega)\Phi_\downarrow(t)
\label{plus1}
\end{equation}

\begin{equation}
\Phi_\downarrow^{\prime\prime}+
\left(\varepsilon_x-t^2\right)\Phi_\downarrow(t)=
(l_\omega/l_\alpha)(k_yl_\omega)\Phi_\uparrow(t).
\label{minus1}
\end{equation}
Unlike the general problem (\ref{plus}) and (\ref{minus}),
Eqs.~(\ref{plus1}) and (\ref{minus1}) can be simply decoupled by a unitary
transformation in the spin space. The corresponding matrix has the form

\begin{equation}
\left(
\begin{array}{cr}
 1/\sqrt{2} & -1/\sqrt{2} \\
 1/\sqrt{2} & 1/\sqrt{2}
\end{array}
\right).
\label{S}
\end{equation}
It is straightforward to verify that eigenvalues (energies) of the
decoupled equations are

\begin{equation}
\varepsilon_x^{\uparrow\downarrow} = 2n+1 \mp
(l_\omega/l_\alpha)(k_yl_\omega)
\label{eps-large}
\end{equation}
and eigenfunctions are identical to those given by Eq.~(\ref{phi0}) at
$l_\omega/l_\beta=0$. The formula (\ref{eps-large}) yields the expected
linear dependence of both $\varepsilon_x^\uparrow$ and
$\varepsilon_x^\downarrow$ on $k_yl_\omega$ with slopes $\mp
(l_\omega/l_\alpha)$ independent of a level number $n$. The asymptotics
(\ref{eps-large}) give a very accurate fit to the spectral pattern in the
region $k_yl_\omega \gtrsim 5$ in Fig.~\ref{fig3}(a). Since
Eqs.~(\ref{plus1}) and (\ref{minus1}) can be decoupled by a simple
rotation, states $\Phi_\uparrow^n$ and $\Phi_\downarrow^n$ turn out to be
intrinsically independent of each other as well as of all the other states.
This fact alone explains why the anticrossing of energy branches is not
observed for sufficiently large values of $k_yl_\omega$.

Let us now consider how the energy spectrum is modified by switching on the
$\beta$-coupling [Fig.~\ref{fig3}(b)]. From visual comparison of
Figs.~\ref{fig3}(a) and (b) we see that the main effect of the
$\beta$-coupling is to {\it enhance} considerably the anticrossing of
``neighbouring'' spectrum branches. Moreover, the strength of the
anticrossing now depends on the quantum number $n$ and grows as $n$ is
increased. An interesting consequence of this behaviour is an {\it
essential reduction} of the linear Rashba energy splitting
$\Delta_R \propto k_y$, contrasting the expectation that the additional
mechanism of the SO interaction should intensify the splitting rather than
suppress it.

To understand the peculiarities of Fig.~\ref{fig3}(b) we note that in
subsection~\ref{subsec-2B} it was shown that a finite $\beta$-coupling
leads to spatial separation between ``up'' and ``down'' spinor components
(see Fig.~\ref{fig1}). As a result, the orthogonality condition
(\ref{norm2}) no longer applies. Instead, the asymptotics (\ref{nn}) --
(\ref{nn+p}) should be used. If we now calculate the Rashba splitting
$\Delta_R$ (\ref{DR0}), then we will see that scalar products
$\left<\phi_{\uparrow\downarrow}^n | \phi_{\downarrow\uparrow}^{n \pm
1}\right>$ give non-zero (viz linear in $l_\omega/l_\beta$) contributions
to the diagonal matrix elements $P_{\uparrow\downarrow}^{nn}$, so that
$\Delta_R$ is now described by [cf. Eq.~(\ref{DR1})]

\begin{equation}
\Delta_R \simeq
2\left[1-\left(n+\frac{1}{2}\right)\frac{l_\omega}{l_\beta}\right]
\frac{l_\omega}{l_\alpha}(k_yl_\omega).
\label{DR2}
\end{equation}
From this formula it immediately follows that the energy splitting
$\Delta_R$ {\it diminishes} in comparison with Fig.~\ref{fig3}(a) by an
amount $(n+1/2)(l_\omega/l_\beta)$ that is {\it proportional} to the
quantum level number $n$. This conclusion agrees well with data presented
in Fig.~\ref{fig3}(b).

The result (\ref{DR2}) can be interpreted in the language of interstate
coupling. Actually, it is easy to verify that the asymptotics (\ref{nn}) --
(\ref{nn+p}) give rise to an additional (proportional to
$l_\omega/l_\beta$) term $\delta\Phi_\beta^n(t)$ in the wave function
correction $\delta\Phi_{\uparrow\downarrow}^n(t)$ (\ref{dphi0}):

\begin{eqnarray}
\delta\Phi_\beta^n(t) \simeq
\frac{1}{2}\frac{l_\omega}{l_\alpha}\frac{l_\omega}{l_\beta}
(k_yl_\omega)\Biggl[&\pm&\sqrt{\frac{n+1}{2}}(k_yl_\omega)
\phi_{\uparrow\downarrow}^{n+1}(t) \nonumber \\
&\mp& \sqrt{\frac{n}{2}}(k_yl_\omega)
\phi_{\uparrow\downarrow}^{n-1}(t)  \nonumber \\
+&\frac{1}{4}&\sqrt{(n+1)(n+2)}
\phi_{\uparrow\downarrow}^{n+2}(t) \nonumber \\
-&\frac{1}{4}&\sqrt{n(n-1)}
\phi_{\uparrow\downarrow}^{n-2}(t)\Biggr]
\label{dphib}
\end{eqnarray}
This term, as well as $\delta\Phi_{\uparrow\downarrow}^n$ (\ref{dphi0}),
involves ``nearest'' unperturbed wave functions
$\phi_{\uparrow\downarrow}^{n \pm 1}$. Since parameters
$l_\omega/l_\alpha$ and $l_\omega/l_\beta$ are independent of each other,
we conclude that the $\beta$-coupling enhances the hybridisation between an
$n$-th and $(n \pm 1)$-st electron states and thereby the anticrossing of
corresponding energy branches. The strength of the hybridisation grows with
the growth of $n$.

Because of the enhanced interstate coupling, the anticrossing of
neighbouring energy branches in Fig.~\ref{fig3}(b) is not restricted to a
narrow region of relatively small values of $k_yl_\omega \lesssim 5$ as it
was in Fig.~\ref{fig3}(a). Instead, we can see that for all $n>0$ there
exists a second anticrossing in the region $k_yl_\omega \approx 7 - 10$.
Moreover, for larger quantum numbers $n \geq 4$ a third anticrossing
emerges at $k_yl_\omega \approx 13 - 14$.

\subsection{Ballistic conductance}
\label{subsec-2D}

To apply the results of the previous subsections to the study of the
effect of the SO interaction on the ballistic conductance of a long Q1DES
at low temperature, we must relate its conductance to its energy spectrum.
Here we do this using the two-probe Landauer formula~\cite{Landauer}

\begin{equation}
G \equiv G(\varepsilon_F) = \frac{e^2}{h}M(\varepsilon_F),
\label{G}
\end{equation}
where $G$ is the ballistic conductance, $\varepsilon_F$ is the Fermi
energy, and $M(\varepsilon_F)$ is the number of occupied electron
subbands which propagate in the same direction:

\begin{equation}
M(\varepsilon_F) =
\sum_n\sum_i\sum_{s=\uparrow,\downarrow}\theta\left[
\varepsilon_F-\varepsilon_{min}^s(n,i)\right].
\label{M}
\end{equation}
Here $\varepsilon_{min}^s(n,i)$ is the energy of the $i$-th minimum in the
$n$-th electron subband with the spin orientation $s$. $\theta(x)$ is the
Heaviside unit step function. As will be seen later the SO interaction can
produce multiple minima in each 1D subband owing to the anticrossings
between different subbands. Since $\varepsilon_{min}^s(n,i)$ can be found
directly from the dispersion law of electrons, the conductance $G$
(\ref{G}) turns out to be {\it completely defined} by the energy spectrum
alone. Therefore, as long as the expression (\ref{G}) applies, the
knowledge of the electron energy levels in a Q1DES will be sufficient to
predict the behaviour of the ballistic conductance as a function of the
Fermi energy.

The applicability of Eq.~(\ref{G}), as well as of the general scattering
approach~\cite{Landauer} to quantum transport, is essentially
based on the condition that a current must travel in any 1D electron
subband without scattering into any other. In Ref.~\cite{Baranger} it was
shown that this {\it current conservation} condition holds true in quite
general circumstances including placing a Q1DES in both a finite external
magnetic field and an arbitrary external electrostatic potential. However,
the proof in Ref.~\cite{Baranger} ignored spin degrees of freedom and
therefore did not take into account potentials acting in the spin space.
Such neglect is not valid for our problem where the SO interaction
Hamiltonian (\ref{H_SO}) leads to a highly non-trivial role for the spin in
forming the energy spectrum of electrons (see subsections~\ref{subsec-2A}
-- \ref{subsec-2C}). For that reason we cannot rely on the conclusion of
Ref.~\cite{Baranger} but should check explicitly if the current is still
conserved in the presence of a finite SO coupling.

The first step is to define the matrix elements of the current density
${\bf j}_{mn}({\bf r})$  for the case where the wave functions
$\Psi_{m(n)}({\bf r})$ are spinors~\cite{Thankappan}, $\Psi_{m(n)} =
\left\{\Psi_\uparrow^{m(n)}, \Psi_\downarrow^{m(n)}\right\}$:

\begin{equation}
{\bf j}_{mn}({\bf r})=\frac{1}{2M}\Biggl\{\Psi_m^\dagger\;
\hat{\bbox{\sigma}}\left(\hat{\bbox{\sigma}}\cdot\hat{\bf p}
\right)\Psi_n+\left[\left(\hat{\bbox{\sigma}}\cdot\hat{\bf p}\right)
\Psi_m\right]^\dagger\hat{\bbox{\sigma}}\Psi_n\Biggr\}.
\label{jmn}
\end{equation}
Here the dagger denotes the hermitian conjugate and
$\hat{\bf p}=-i\hbar{\bf \nabla}$. It is straightforward to verify that the
divergence of the vector ${\bf j}_{mn}({\bf r})$ is given by

\begin{equation}
{\bf \nabla}\cdot{\bf j}_{mn}({\bf r})=\frac{i\hbar}{2M}\Biggl\{
\left(\nabla^2\Psi_m^\dagger\right)\Psi_n-
\Psi_m^\dagger\left(\nabla^2\Psi_n\right)\Biggr\}.
\label{dj}
\end{equation}
We now suppose that the Hamiltonian of the system has the general form
$\hat H=\hat{\bf p}^2/2M+\hat Q$, where $\hat Q$ is a hermitian operator
(${\hat Q}^\dagger=\hat Q$) that acts in both coordinate and spin
spaces. As applied to our problem, $\hat Q=V_{LC}+\hat H_{SO}$ [see
Eq.~(\ref{H})]. Using this Hamiltonian and the Schr\"{o}dinger equation
$\hat H\Psi=E\Psi$, we express $\nabla^2\Psi_m^\dagger$ and
$\nabla^2\Psi_n$ in terms of $\Psi_m^\dagger$ and $\Psi_n$ respectively and
substitute the expressions obtained into Eq.~(\ref{dj}). As a result, we
have

\begin{equation}
{\bf \nabla}\cdot{\bf j}_{mn}({\bf r})=\frac{i}{\hbar}(E_m-E_n)
\Psi_m^\dagger({\bf r})\Psi_n({\bf r}),
\label{dj2}
\end{equation}
where $E_m$ and $E_n$ are energies corresponding to the states $\Psi_m$ and
$\Psi_n$ respectively. In the absence of any inelastic collisions, any
scattering occurs between states of the same energy. So, without loss of
generality, we can restrict ourselves to considering only equal energies
$E_m=E_n$ in Eq.~(\ref{dj2}), in which case we find that

\begin{equation}
{\bf \nabla}\cdot{\bf j}_{mn}({\bf r})=0
\qquad\mbox{for}\qquad E_m=E_n.
\label{dj=0}
\end{equation}
This fundamental identity ensures {\it local} current conservation in
the system. Once it has been established, if the system is translationally
invariant in the longitudinal direction, further steps in proving the {\it
global} conservation of current do not depend on specifics of the
Hamiltonian and can be carried out in line with Ref.~\cite{Baranger}.
Therefore we arrive at the conclusion that the total current (i.e.
current integrated over the cross section of the channel) between states
$m$ and $n$ at $E_m=E_n$ is equal to zero unless $m=n$. In other words,
eigenstates of the Hamiltonian $\hat H$ are perfect current-carrying states
that are free from scattering even in the presence of arbitrary SO coupling.

For our problem this result implies that the spectrum of the Hamiltonian
(\ref{H}) is directly relevant to and completely defines the ballistic
conductance in the presence of the SO coupling. This allows us to use the
simple Landauer formula (\ref{G}), (\ref{M}), in which the minima
$\varepsilon_{min}^s(n,i)$ of the energy subbands can be found from the
analysis of the spectrum presented in subsections~\ref{subsec-2A} --
\ref{subsec-2C}.

Now we are in a position to discuss the features of the ballistic
conductance in a Q1DES subject to the SO interaction. For illustrative
purposes we start with the ``ideal'' case of zero SO coupling. The
corresponding subband energies $\varepsilon(n)=2n+1+(k_yl_\omega)^2$ are
plotted in Fig.~\ref{fig5}(a) (solid curves) as functions of $k_yl_\omega$.
The dependence $G(\varepsilon_F)$ can simply be deduced from this figure by
moving a horizontal line $\varepsilon=\varepsilon_F$ from zero upwards and
counting the number of points at which this line crosses the spectral
parabolas. Since all the subbands in an ideal system are two-fold spin
degenerate for any $k_y$, this number coincides with the number $M$ of
propagating modes in the Q1DES. As a result, we restore the well-known
picture~\cite{BvH,Kelly,vanWees,Wharam,Datta} of ballistic conductance
quantisation with equidistant jumps each of height $2e^2/h$
[solid curve in Fig.~\ref{fig6}(a)].

\vspace{0.5cm}
\begin{figure}
\begin{center}
\epsfig{file=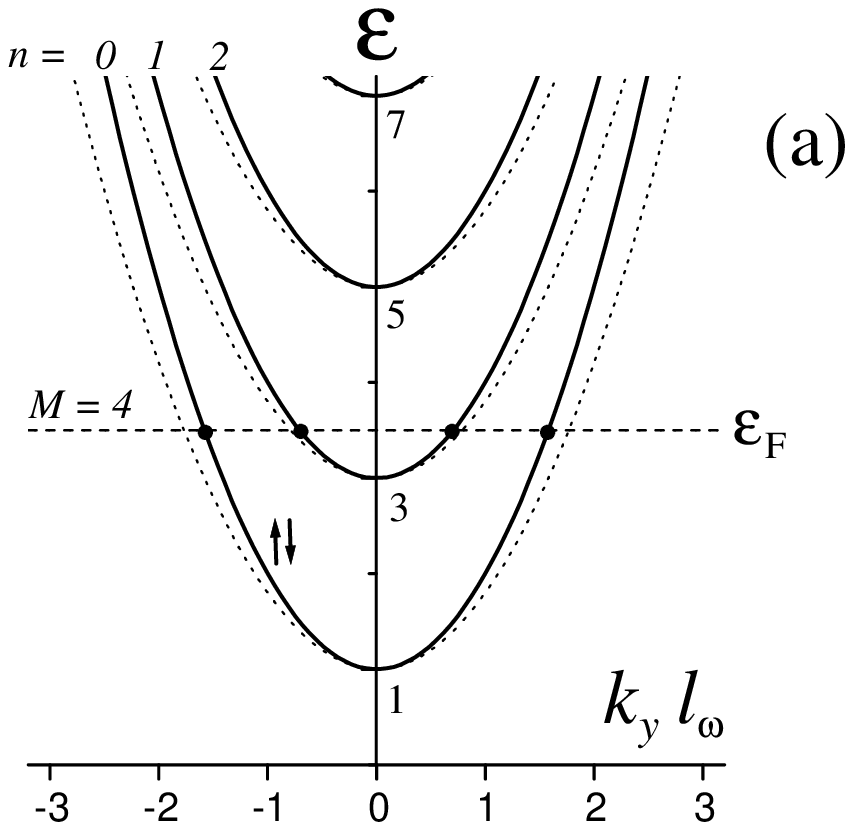,width=3.00in}
\end{center}
\end{figure}

\begin{figure}
\begin{center}
\epsfig{file=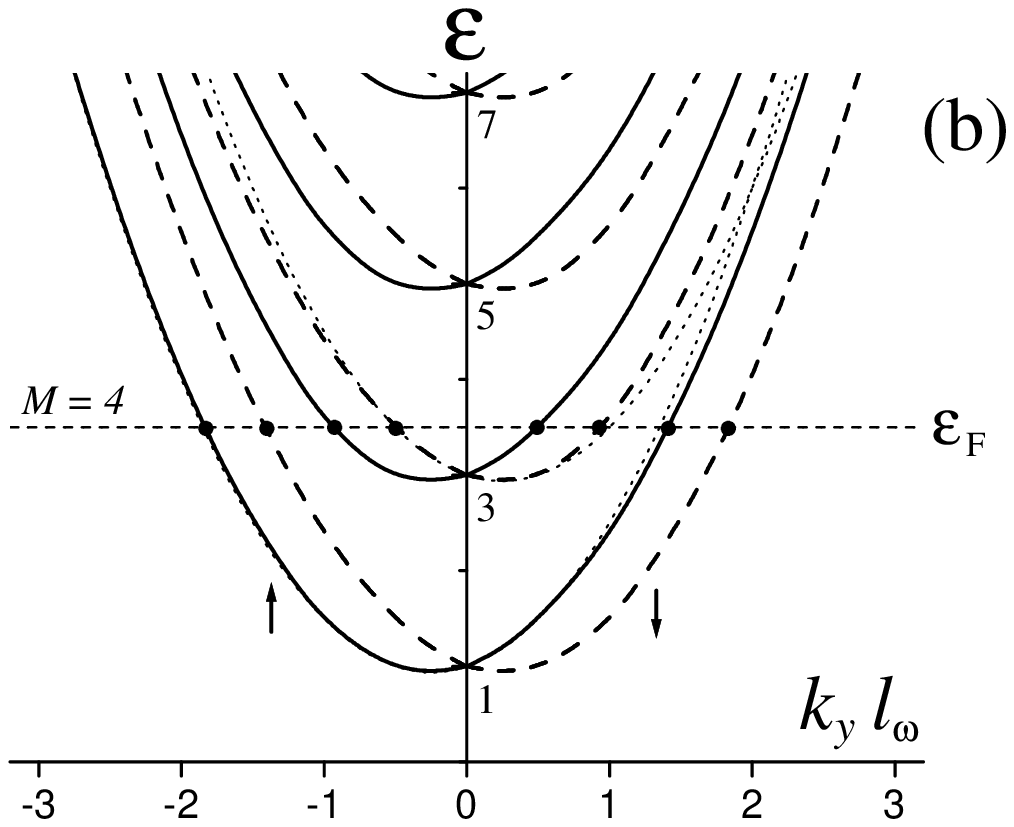,width=3.00in}
\end{center}
\end{figure}

\begin{figure}
\begin{center}
\epsfig{file=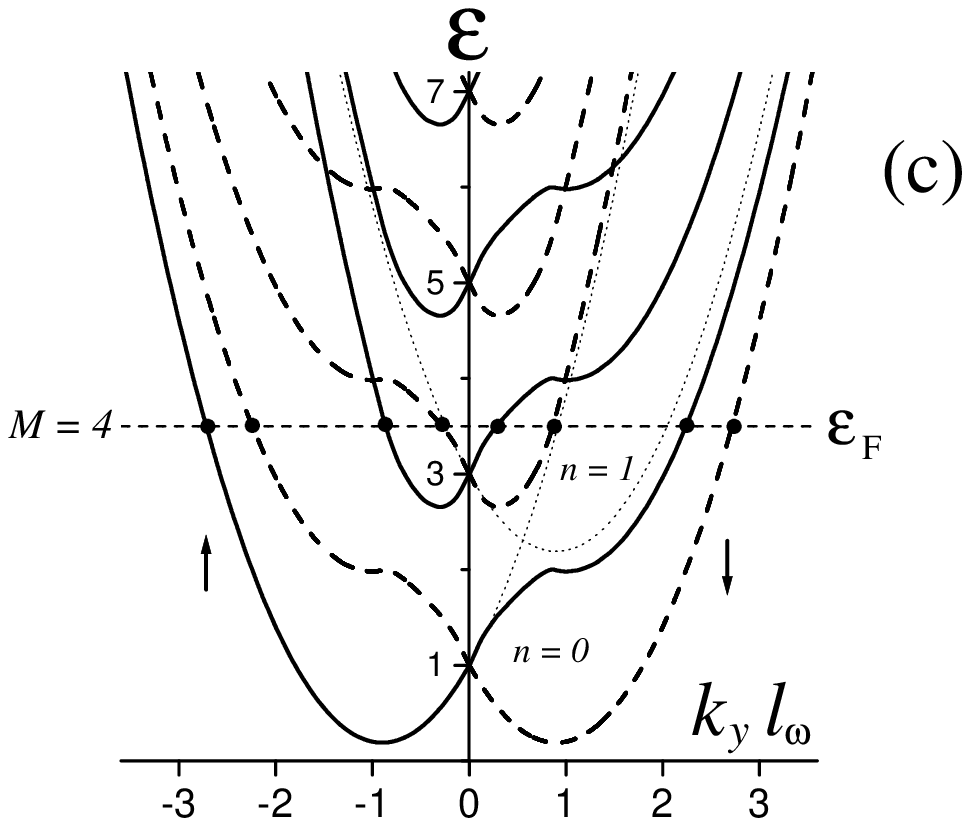,width=3.00in}
\end{center}
\end{figure}

\begin{figure}
\begin{center}
\epsfig{file=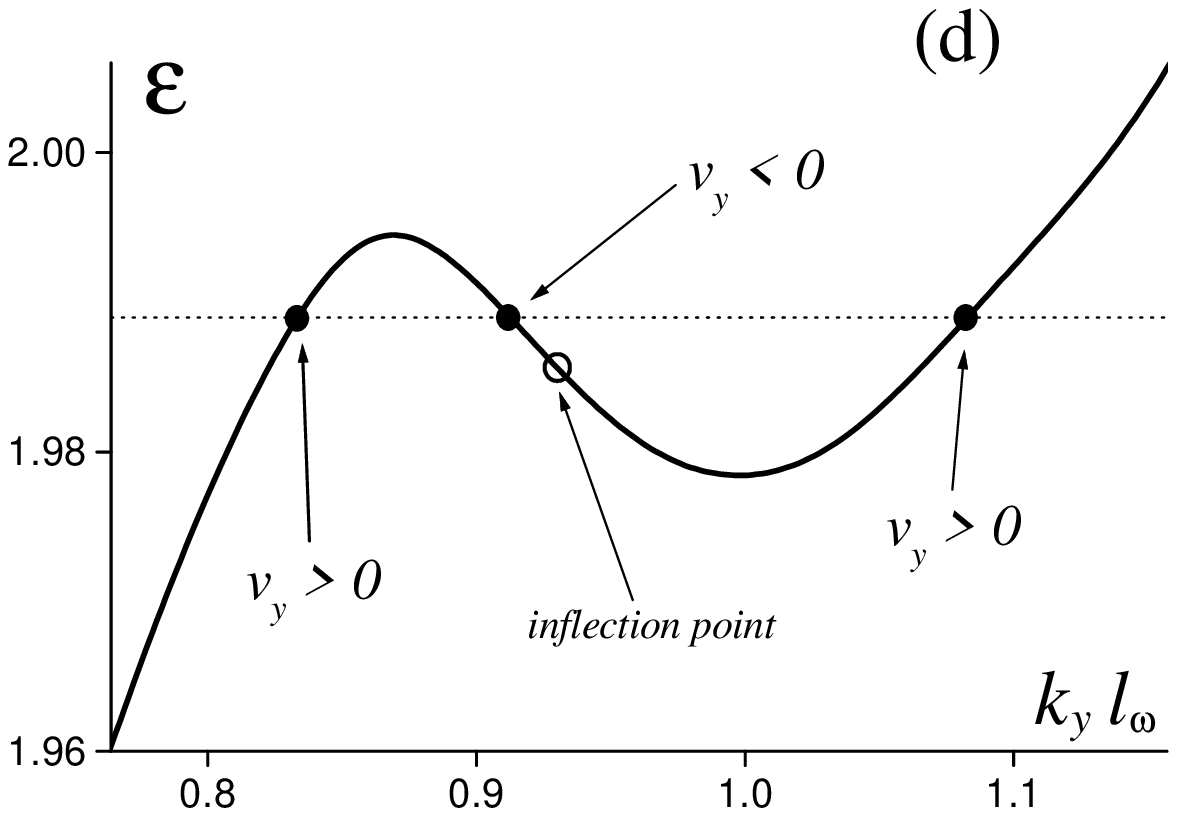,width=3.00in}
\end{center}
\end{figure}

\begin{figure}
\begin{center}
\epsfig{file=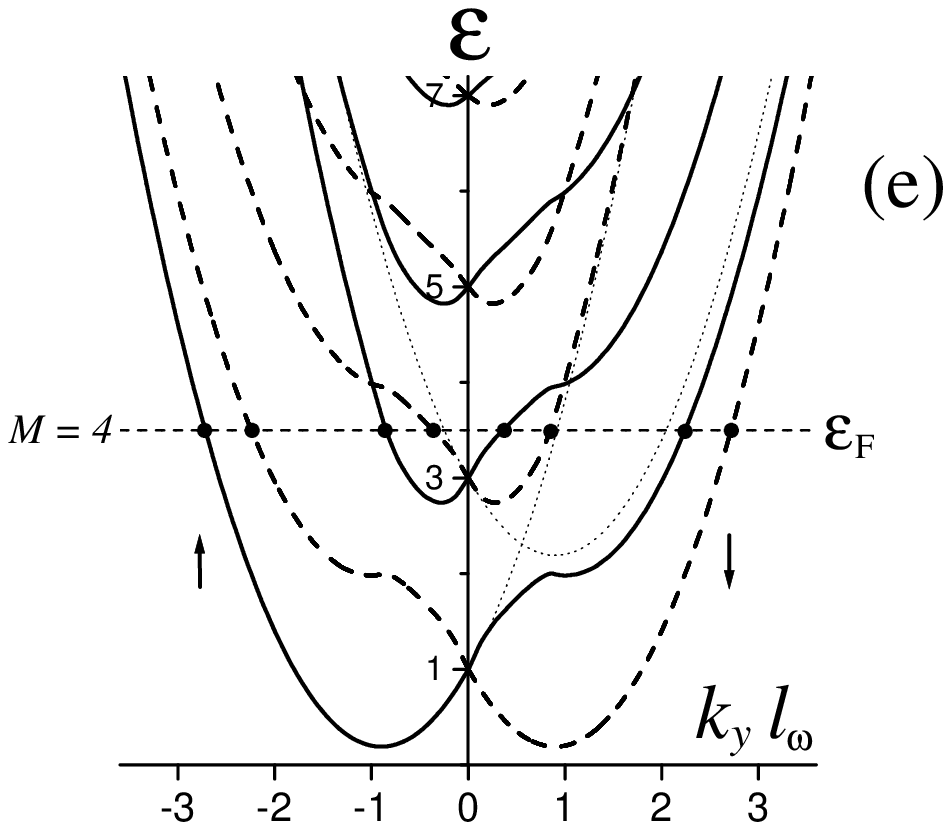,width=3.00in}
\caption{The subband energy $\varepsilon$ in units of $\hbar\omega/2$ vs.
$k_yl_\omega$: (a) zero SO coupling (solid curves), finite
$\beta$-coupling (dotted curves); (b) weak $\alpha$-coupling
($l_\omega/l_\alpha < \sqrt{2}$); (c) strong $\alpha$-coupling
($l_\omega/l_\alpha > \sqrt{2}$) and zero $\beta$-coupling; (d) a magnified
bump on an $n=0$ energy branch in the anticrossing region; (e) strong
$\alpha$-coupling ($l_\omega/l_\alpha > \sqrt{2}$) and finite
$\beta$-coupling.}
\label{fig5}
\end{center}
\end{figure}

\vspace{0.5cm}
\begin{figure}
\begin{center}
\epsfig{file=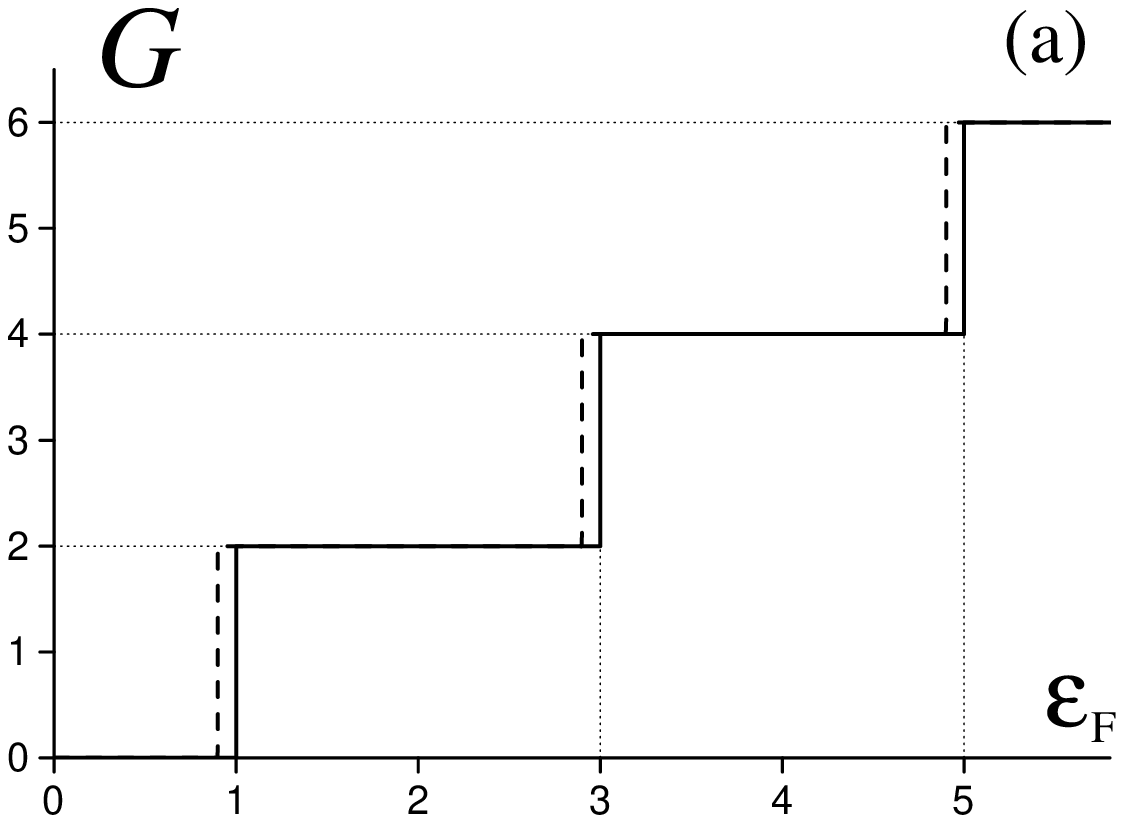,width=3.00in}
\end{center}
\end{figure}

\begin{figure}
\begin{center}
\epsfig{file=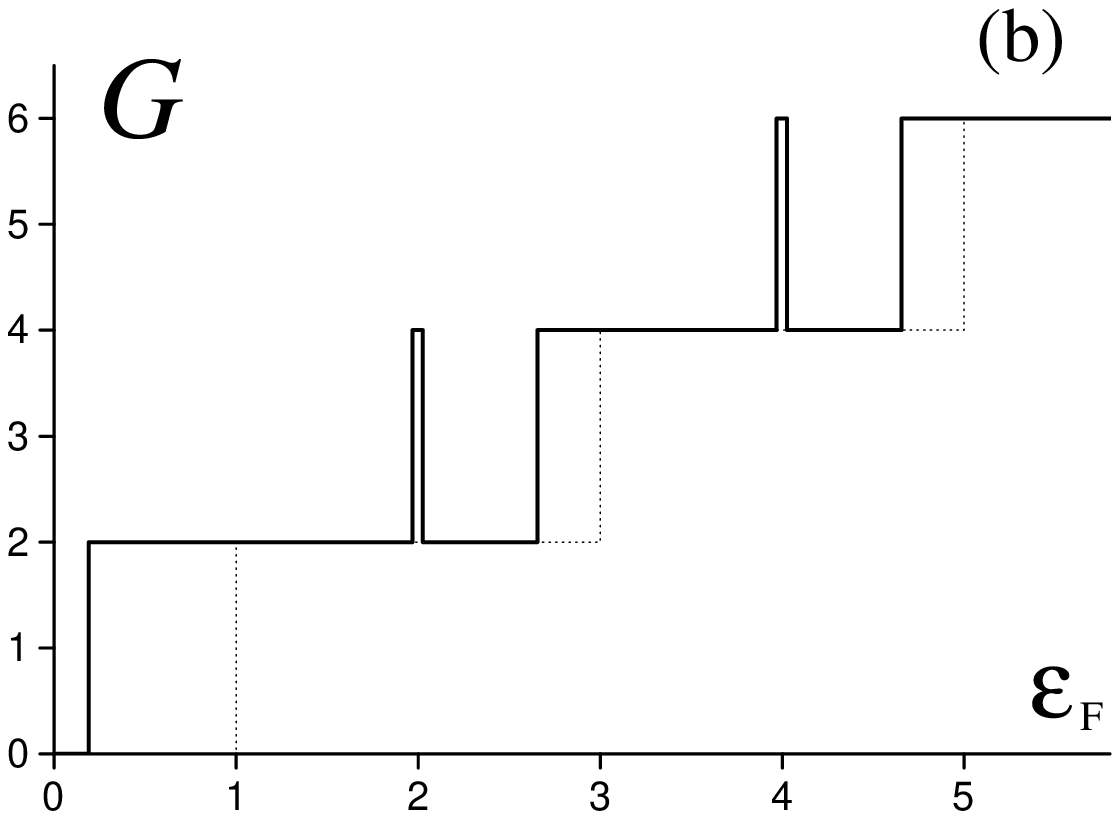,width=3.00in}
\end{center}
\end{figure}

\begin{figure}
\begin{center}
\epsfig{file=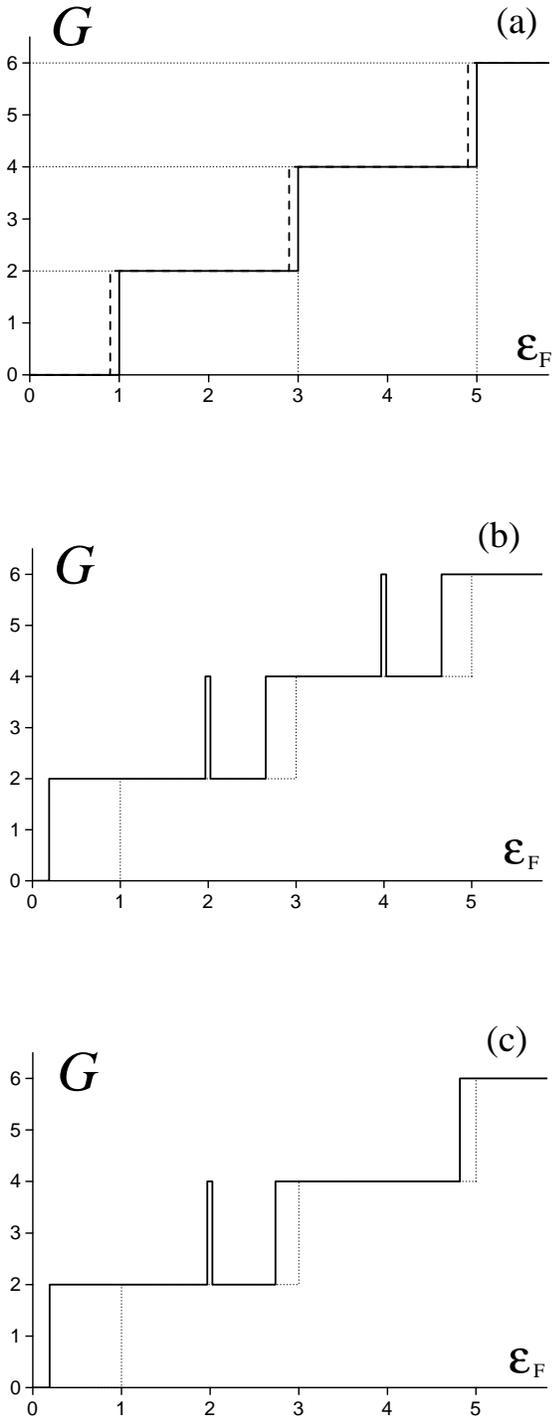,width=3.00in}
\caption{The conductance $G$ in units of $e^2/h$ vs. the Fermi energy
$\varepsilon_F$ in units of $\hbar\omega/2$: (a) zero SO coupling and
finite $\beta$-coupling (solid line), weak $\alpha$-coupling (dashed line);
(b) strong $\alpha$-coupling and zero $\beta$-coupling; (c) strong
$\alpha$-coupling and finite $\beta$-coupling.}
\label{fig6}
\end{center}
\end{figure}

Dotted curves in Fig.~\ref{fig5}(a) show subband energies
$\varepsilon(n)=\varepsilon_x^{(0)}(n)+(k_yl_\omega)^2$ in the presence of
the $\beta$-coupling [see Eq.~(\ref{spec0})]. Each subband is two-fold spin
degenerate. It is seen that for any given value of $\varepsilon_F$ the
number $M$ of forward propagating modes remains the same as it was in the
ideal limit. Thus, we conclude that the $\beta$-coupling alone does {\it
not} affect the ballistic conductance. To detect the presence of the
$\beta$-coupling in a system one should exploit a unique feature that can
only be due to the $\beta$-coupling. As a reasonable example of such a
feature, we mention the spatial separation between spinor components that
was established in subsection~\ref{subsec-2B} and illustrated in
Fig.~\ref{fig1}. Owing to this effect, the regions of the most probable
distribution of electrons with opposite spin orientations will be shifted
apart in the transverse direction. By analogy with ferromagnetic materials,
this effect can manifest itself in the appearance of the magnetisation of
the electron system and its dependence on the transverse coordinate.

Now we move on to the more complex case of finite $\alpha$-coupling. To
start with let us imagine the fictitious situation where the linear Rashba
energy splitting $\Delta_R=2(l_\omega/l_\alpha)(k_yl_\omega)$ [see
Eq.~(\ref{DR1}) and Fig.~\ref{fig3}(a)] holds true not only for small
values of $k_yl_\omega$ but for all $k_yl_\omega$. Here, the subband
energies would be given by
$\tilde\varepsilon_{\uparrow\downarrow}(n)=2n+1+(k_yl_\omega)^2 \pm
(l_\omega/l_\alpha)(k_yl_\omega)$. This formula describes parabolas of the
same form as the solid curves in Fig.~\ref{fig5}(a) but shifted by an
amount $\pm (1/2)(l_\omega/l_\alpha)$ along the abscissa and lowered
vertically by $(1/4)(l_\omega/l_\alpha)^2$. The parabolas
$\tilde\varepsilon_{\uparrow\downarrow}(n)$ were used in Ref.~\cite{Fasol}
to predict spontaneous spin polarisation of ballistic electrons in quantum
wires due to spin splitting. Whereas the {\it real} energy branches
$\varepsilon_\uparrow(n)$ and $\varepsilon_\downarrow(n+1)$ anticross [see
Fig.~\ref{fig3}(a)], the fictitious dispersion curves
$\tilde\varepsilon_\uparrow(n)$ and $\tilde\varepsilon_\downarrow(n+1)$
{\it cross} at the point $k_yl_\omega=(l_\omega/l_\alpha)^{-1}$. This point
remains the same for all numbers $n$ and goes to infinity as the SO
coupling vanishes. It is easy to verify that for $l_\omega/l_\alpha <
\sqrt{2}$ the crossing point $k_yl_\omega=(l_\omega/l_\alpha)^{-1}$ lies to
the right of the point $k_yl_\omega=(1/2)(l_\omega/l_\alpha)$ at which the
parabola $\tilde\varepsilon_\downarrow(n+1)$ has the minimum. At the same
time, if $l_\omega/l_\alpha > \sqrt{2}$, then the crossing point turns out
to be to the left of the minimum point. Below we will see that these simple
conclusions following from the {\it fictitious} energy spectrum play an
important role in determining the behaviour of the conductance
$G(\varepsilon_F)$.

Fig.~\ref{fig5}(b) presents the energy subbands
$\varepsilon_{\uparrow\downarrow}(n)$ (solid and dashed curves) of a Q1DES
in the case of {\it weak} $\alpha$-coupling when $l_\omega/l_\alpha <
\sqrt{2}$ (more specifically, $l_\omega/l_\alpha=0.5$). Dotted lines
indicate the {\it fictitious} energy levels
$\tilde\varepsilon_\uparrow(n=0)$ and $\tilde\varepsilon_\downarrow(n=1)$.
In full accordance with the above conclusions, these levels cross to the
{\it right} of the bottom of the parabola
$\tilde\varepsilon_\downarrow(n=1)$. As a result, the crossing angle turns
out to be quite small. From Fig.~\ref{fig5}(b) it is seen that this angle
determines essentially the shape of the {\it real} energy curves
$\varepsilon_\uparrow(n=0)$ and $\varepsilon_\downarrow(n=1)$ within the
anticrossing region $1.5 \lesssim k_yl_\omega \lesssim 3$. As long as the
angle is small, the anticrossing remains very smooth and the curves
$\varepsilon_\uparrow(n=0)$ and $\varepsilon_\downarrow(n=1)$ behave very
much like the ideal parabolas in Fig.~\ref{fig5}(a). The same observation
is also true for higher quantum numbers $n$. Such similarity suggests that
weak $\alpha$-coupling ($l_\omega/l_\alpha < \sqrt{2}$) does not have a
strong effect on the conductance. Indeed, it is easy to see by scanning
Fig.~\ref{fig5}(b) with the horizontal line $\varepsilon=\varepsilon_F$
that the only effect of  weak $\alpha$-coupling on $G(\varepsilon_F)$ is
to shift the conductance quantisation steps down to lower energies by an
amount  $(1/4)(l_\omega/l_\alpha)^2$ [see dashed lines in
Fig.~\ref{fig6}(a)].

As soon as the coupling constant $l_\omega/l_\alpha$ gets over the
threshold of $\sqrt{2}$, the spectral picture becomes much more interesting.
Fig.~\ref{fig5}(c) shows a case when $l_\omega/l_\alpha=1.8$. Here the
parabolas $\tilde\varepsilon_\uparrow(n=0)$ and
$\tilde\varepsilon_\downarrow(n=1)$ (dotted curves) cross to the {\it left}
of the bottom of $\tilde\varepsilon_\downarrow(n+1)$, which makes the
crossing angle relatively large. The direct consequence of this is that the
energy branches $\varepsilon_\uparrow(n=0)$ and
$\varepsilon_\downarrow(n=1)$ cannot now anticross as smoothly as they did
for $l_\omega/l_\alpha < \sqrt{2}$. Instead, in order to keep continuity,
the lower curve [i.e. $\varepsilon_\uparrow(n=0)$] is forced to exhibit a
non-monotonic portion (``bump'') within the anticrossing region. Although
the height of the bump is rather small ($\sim 0.05$ in the dimensionless
units or $\sim 0.25$ meV as determined from typical subband
spacings~\cite{Koester}), its geometry is a unique feature of the SO
coupling mechanism [see a magnified plot of the bump in
Fig.~\ref{fig5}(d)]. The remarkable fact about it is that the bump contains
a region where the electron energy decreases as the wavenumber grows. In
other words, there exists an energy interval within which electrons are
allowed to have a negative longitudinal group velocity
$v_y=\hbar^{-1}(\partial\varepsilon/\partial k_y) < 0$ for positive $k_y$.
Since this interval has a finite width and is surrounded by energy domains
with positive group velocities, the curvature of the line
$\varepsilon=\varepsilon(k_y)$ reverses sign within the negative-velocity
interval and therefore there exists an inflection point where
$\partial^2\varepsilon/\partial k_y^2=0$ [see Fig.~\ref{fig5}(d)]. At this
point the effective electron mass
$M=\hbar^2\left(\partial^2\varepsilon/\partial k_y^2\right)^{-1}$ diverges.
Since the current passing through a 1D subband is proportional to the
product of the group velocity and the density of states~\cite{Kelly,Datta},
the singular effective mass has no effect on the conductance in the absence
of electron scattering. However, it could significantly change the
dependence $G(\varepsilon_F)$ in the presence of a scattering mechanism
(e.g., disorder or electron-electron interaction).

As far as transport through 1D electron subbands is concerned, it is clear
from Fig.~\ref{fig5}(d) that the negative propagating mode coexists with a
forward propagating mode with the same spatial wave function. In this
physical situation it is likely that weak elastic scattering between these
states would result in directed localisation Ref.~\cite{Barnes} so that
they would not be observed in conductance.

However, if both forward and backward electron modes contribute equally to
the conductance, then this immediately brings us to the conclusion that the
bump on the curve $\varepsilon_\uparrow(n=0)$ gives rise to a peak in the
dependence $G(\varepsilon_F)$ [see Fig.~\ref{fig6}(b)]. The height of the
peak is $2e^2/h$ and its width is defined by the height of the bump.
Analogous bumps can be observed for all quantum levels $n=0,1,2,\ldots$ As
a consequence, the sharp peak is seen in Fig.~\ref{fig6}(b) not only on the
first conductance plateau $G=2e^2/h$ but also on the second one $G=4e^2/h$
and would still be seen on all further plateaus.

Of course, the strong $\alpha$-coupling ($l_\omega/l_\alpha > \sqrt{2}$),
as well as the weak coupling ($l_\omega/l_\alpha > \sqrt{2}$), shifts the
conductance quantisation steps to lower values of the Fermi energy in
comparison with the ideal situation where $l_\omega/l_\alpha=0$. However, in
contrast to the weak-coupling limit, strong coupling makes this shift
much larger for the first step than for the others. This effect is clearly
seen in Fig.~\ref{fig6}(b) as opposed to the ideal conductance quantisation
shown by dotted lines. The larger shift of the first step is explained
by the fact that the $n=0$ quantum state does not have a neighbour with the
next lowest quantum number $n$. The energy level
$\varepsilon_\downarrow(n=0)$ is therefore not forced to anticross with any
other (lower lying) energy branch and therefore nothing affects its linear
(Rashba) dependence on $k_y$ [see Fig.~\ref{fig3}(a)].

The final step of our analysis is to consider the case where both strong
$\alpha$-coupling ($l_\omega/l_\alpha=1.8$) and relatively weak
$\beta$-coupling ($l_\omega/l_\beta=0.2$) are present in the system. The
corresponding energy bands are shown in Fig.~\ref{fig5}(e). As we
demonstrated in subsection~\ref{subsec-2C}, finite $\beta$-coupling tends
to suppress the energy splitting caused by the $\alpha$-coupling [cf.
Figs.~\ref{fig3}(a) and \ref{fig3}(b) and see Eq.~(\ref{DR2})].
Effectively, this suppression can be interpreted as the enhancement of the
anticrossing between neighbouring energy branches $\varepsilon_\uparrow(n)$
and $\varepsilon_\downarrow(n+1)$. In Fig.~\ref{fig5}(e) this effect is
seen as a decrease in the gap between energy branches
$\varepsilon_\uparrow(n)$ and $\varepsilon_\downarrow(n)$
($n=1,2,3,\ldots$) in comparison with Fig.~\ref{fig5}(c). Moreover, in
accordance with Eq.~(\ref{DR2}), the gap becomes monotonically smaller
as $n$ grows. The enhanced anticrossing has a drastic effect on the energy
bumps created by the strong $\alpha$-coupling: all the bumps except for the
lowest one are now smoothed away and do not give rise to the two additional
(forward and backward) propagating modes. As a result, the conductance
exhibits only one sharp peak that occurs on the plateau $G=2e^2/h$ [see
Fig.~\ref{fig6}(c)]. A second effect of the enhanced anticrossing on the
conductance is that the conductance quantisation steps (starting from the
second one) are now located closer to the ideal steps than they were for
zero $\beta$-coupling. As each next step appears, its distance from the
corresponding ideal step diminishes until eventually they fuse.

There is another effect of the enhanced energy anticrossing on the
conductance that is not seen in Figs.~\ref{fig6} but could possibly be
detected in the presence of impurities. It is found
experimentally~\cite{Timp,Koester2} that the ballistic conductance
quantisation breaks down as the constriction becomes too long (at least
$\gtrsim 2$ $\mu$m). This effect can be attributed to potential
fluctuations caused by the random distributions of remote impurities.
Zagoskin {\it et al}~\cite{Zagoskin} showed analytically that the
degradation of quantised conductance decreases exponentially with the ratio
of the 1D subband spacings to the standard deviation of impurity potential
fluctuations. Obviously, the subband spacings depend on the presence of the
SO interaction (see Figs.~\ref{fig5}). By comparing Figs.~\ref{fig5}(c) and
(e) we see that a finite $\beta$-coupling pushes neighbouring energy
branches apart and thus increases the subband spacings (possibly by tens of
per cent). This means that the presence of the $\beta$-mechanism of the SO
interaction could allow the experimental observation of the conductance
quantisation structure in longer samples.

%%%%%%%%%%%%%%%%%%%%%%%%%%%%% SECTION %%%%%%%%%%%%%%%%%%%%%%%%%%%%%%%%%%%
\section{Conclusion}
\label{sec-con}
%%%%%%%%%%%%%%%%%%%%%%%%%%%%%%%%%%%%%%%%%%%%%%%%%%%%%%%%%%%%%%%%%%%%%%%%%

We have studied theoretically the influence of the spin-orbit (SO)
interaction on the energy spectrum and the ballistic conductance of a
quasi-1D electron system (Q1DES) formed by lateral electric confinement of
a 2D electron gas. The presence of the confining potential proves to be
crucial in two ways. First, it alters significantly the effect of the
quantum well asymmetry on the energy spectrum in comparison with a purely
2D system. As it was shown by Rashba {\it et al}~\cite{Rashba,Bychkov}, in
2D electron gases this asymmetry gives rise to SO coupling (which we refer
to as the $\alpha$-coupling) that manifests itself in a lifting of the spin
degeneracy of electronic states. The accompanying energy splitting
$\Delta_R$ was found to be proportional to the in-plane electron wave
number $k$. We have demonstrated that in a Q1DES the function $\Delta_R(k)$
is {\it non-monotonic} [Fig.~\ref{fig3}(a)] with the standard linear
dependence~\cite{Rashba,Bychkov} $\Delta_R \propto k$ being observed for
relatively small and large values of $k$ only. Such a drastic change in
behaviour is explained by the essentially different effect of SO coupling
operator $\hat P_{\uparrow\downarrow}$ (\ref{P0}) on unperturbed wave
functions depending on whether or not the lateral confining potential is
taken into account. Indeed, in the zero-potential case the unperturbed wave
functions are plane waves characterised by the continuous in-plane wave
vector ${\bf k}$ and the spin. The action of the operator $\hat
P_{\uparrow\downarrow}$ on any such wave function reduces to a simple
renormalisation of its amplitude with the renormalising factor being
independent of coordinates. As a consequence of this simple effect, wave
functions belonging to different values of ${\bf k}$ remain independent of
each other and the SO coupling manifests itself in linear (monotonic)
energy splitting $\Delta_R \propto k$ which is the same for all quantum
states.

In contrast to this, the wave functions in the presence of a finite
confining potential have a more complicated structure (\ref{phi0}) and are
characterised by the discrete quantum number $n$ instead of the continuous
transverse wave number. According to Eq.~(\ref{deriv-prop}), the action of
the operator $\hat P_{\uparrow\downarrow}$ on the $n$-th wave function
includes projecting it onto states with the next (preceding) closest
quantum numbers $n \pm 1$. In turn, the $(n \pm 1)$-st states are projected
onto their ``neighbours'' with numbers $n$ and $n \pm 2$ respectively. As a
result, the operator $\hat P_{\uparrow\downarrow}$ {\it couples}
effectively any $n$-th and $(n \pm 1)$-st wave functions. In other words,
once the SO coupling has been taken into account, the $n$-th and $(n \pm
1)$-st wave functions cease to be independent and possess symmetry elements
of each other. This partial symmetry between states leads to an {\it
anticrossing} of the closest (neighbouring) energy branches in
Fig.~\ref{fig3}(a) and hence to the non-monotonic dependence $\Delta_R(k)$.

Apart from the interplay with the familiar (quantum-well-asymmetry or
Rashba) mechanism of the SO interaction, the lateral confining potential by
itself appears to be a source of additional dynamical coupling between
the orbital and spin degrees of freedom of an electron. This coupling
(which we refer to as $\beta$-coupling) originates from the natural spatial
non-uniformness of the confining potential. A typical variation scale of
this potential lies within the wide range $\sim 10$ -- $1000$ nm, which
makes the accompanying electric field sufficiently strong to compete with
the quantum-well-asymmetry field. This competition may become especially
noticeable in square quantum wells with a relatively weak Rashba
contribution. Whereas the quantum-well field is normal to the device plane,
the confinement-induced electric field is parallel to the plane. It
is also spatially non-uniform as long as the coordinate dependence of the
confining potential is more complex than linear. These features make the
$\beta$-coupling an essentially different (new) mechanism of the SO
interaction in a Q1DES which {\it cannot} be taken into account by simply
adjusting the Rashba interaction constant. This claim is confirmed by
Fig.~\ref{fig3}(b) that demonstrates the combined effect of both the Rashba
and the confinement-induced couplings on the energy spectrum. From
comparison of Fig.~\ref{fig3}(b) with Fig.~\ref{fig3}(a) we see that the
major role of $\beta$-coupling is to {\it reduce} the Rashba energy
splitting $\Delta_R$ (see region $k_yl_\omega \lesssim 2$) rather than to
give a positive correction to it. It is crucial that the reduction depends
on the quantum number $n$ and monotonically grows as $n$ increases. This
fact is a clear manifestation of the position dependence of the electric
field created by the confining potential and indicates the independent
nature of $\beta$-coupling. The suppression of the Rashba energy splitting
is a part of the overall effect of the $\beta$-coupling on the underlying
electron system which consists of a considerable enhancement of the
hybridisation between different quantum states and eventually leads to a
more pronounced anticrossing of neighbouring energy branches.

In addition to the investigation of the SO-effects in the energy spectrum
of electrons, we have discussed the possible manifestations of the SO
interaction in the ballistic conductance of a Q1DES. The key point of our
approach to the conductance is the fundamental {\it current-conservation
identity} that was proven in subsection~\ref{subsec-2D}. According to this
identity, the electron eigenstates that were found as the solution to the
spectral problem are {\it perfect} current-carrying states. A current can
travel in any of these states without scattering into any other. This
property therefore allows the ballistic conductance to be calculated
directly from the energy spectrum with the help of simple Landauer formula
(\ref{G}).

An analysis of the ballistic conductance $G$ reveals that the
$\beta$-coupling alone does {\it not} affect the dependence of $G$ on the
Fermi energy $\varepsilon_F$. This fact is illustrated by
Fig.~\ref{fig5}(a) where we see that the $\beta$-coupling reduces the
curvature of the parabolic energy bands (cf. dotted and solid curves),
while the band edges (bottoms of the parabolas) remain anchored. Thus,
$\beta$-coupling cannot be detected by measuring the ballistic conductance
in systems where the SO interaction appears {\it only} in a form of the
$\beta$-coupling (e.g. in square wells). Instead, one should use
experimental methods that allow direct observation of the electron energy
spectrum in a Q1DES, e.g. magnetotunneling measurements~\cite{Kardynal}.

In the presence of $\alpha$-coupling the behaviour of $G(\varepsilon_F)$ is
essentially determined by the strength of the SO interaction. If the
$\alpha$-coupling is not too strong ($l_\omega/l_\alpha < \sqrt{2}$), then
its only effect on the conductance will be shifts of the conductance
quantisation steps to lower Fermi energies in comparison with an ideal
(i.e. with zero SO interaction) situation. We note that such shifts are not
related to the lateral confining potential and should also be present in
purely 2D systems. This effect should be detectable in transconductance
measurements (see, e.g., Ref.~\cite{Thomas}) which determine both the
conductance and the subband spacings simultaneously.

In the limit of strong $\alpha$-coupling ($l_\omega/l_\alpha > \sqrt{2}$)
the electron energy bands take on a very interesting form
[Fig.~\ref{fig5}(c)]. The most remarkable feature is the appearance of
narrow energy intervals where two additional (forward and backward
propagating) electron modes exist [Fig.~\ref{fig5}(d)]. Such intervals can
be found in each 1D subband, starting from the lowest one. The additional
electron modes have similar magnitude group velocities but propagate in
opposite directions. They have almost identical subband wave functions and
therefore would be susceptible to strong intermode scattering in the
presence of disorder. However, in a sufficiently pure Q1DES, the additional
electron modes give rise to the unusual periodic sharp steps in
$G(\varepsilon_F)$ shown in Fig.~\ref{fig6}(b). This picture is changed by
switching on relatively weak ($l_\omega/l_\beta \ll l_\omega/l_\alpha$)
$\beta$-coupling. As we mentioned above in the discussion of the energy
spectrum, the $\beta$-coupling enhances the anticrossing of energy levels
initiated by the $\alpha$-coupling. As applied to the conductance, this
effect leads to {\it quenching} the sharp conductance peaks by the
$\beta$-coupling. The existence of the single peak (or just a few of peaks)
in the dependence of $G$ on $\varepsilon_F$ could be a clear experimental
indication of the presence of the $\beta$-coupling in the system.

The most crucial point for the experimental observation of the conductance
peaks in Fig.~\ref{fig6} is to make the ratio $l_\omega/l_\alpha$
sufficiently large ($l_\omega/l_\alpha>\sqrt{2}$). In typical systems,
where enhancing the SO interaction is not paid special attention, the value
of $l_\omega/l_\alpha$ hardly exceeds $0.5$. An additional, at least
threefold, increase in $l_\omega/l_\alpha$ could be achieved by using: (i)
materials with light carrier masses; (ii) strong (narrow) lateral potential
confinement $\lesssim 100$ nm; (iii) heterojunctions (triangular well)
rather than quantum-well heterostructures (square well); (iv) a back gate
voltage to maximise the interface (Rashba) electric field.

We hope that the results presented in this paper will stimulate further
experimental and theoretical work with the aim of understanding the role of
the spin-orbit interaction in determining the transport properties of
quasi-1D systems.

\acknowledgments

AVM thanks the ORS, Cambridge Overseas Trust and Corpus Christi College for
the Research Studentship. CHWB thanks the EPSRC for the Advanced
Fellowship. This work was funded by the EPSRC.

%************************** BIBLIOGRAPHY ********************************

\end{multicols}
\end{document}